\documentclass[11pt]{article}
\newcommand{\Comment}[1]{{}}
\usepackage{amsfonts,amsthm,amsmath,amssymb,slashed}
\usepackage[textwidth = 430 pt, textheight = 630 pt]{geometry}
\usepackage{color}

\Comment{\usepackage{color}
\definecolor{MyDarkBlue}{rgb}{0.15,0.15,0.45}
\usepackage[linktocpage=true]{hyperref}
\hypersetup{
colorlinks=true,
citecolor=MyDarkBlue,
linkcolor=MyDarkBlue,
urlcolor=MyDarkBlue,
pdfauthor={Horatiu Nastase and Jacob Sonnenschein},
pdftitle={The title},
pdfsubject={hep-th}
}

\usepackage[numbers,sort&compress]{natbib}
\usepackage{hypernat}}
\usepackage{graphicx}
\usepackage{cite}
\usepackage{url}

\newcommand\ignore[1]{}
\def\one{{\,\hbox{1\kern-.8mm l}}}

\def\Tr{{\rm Tr\, }}

\def\a{\alpha}\def\b{\beta}

\def\d{\partial}

\def\Tr{\mathop{\rm Tr}\nolimits}

\newcommand{\Cset}{{\,\,{{{^{_{\pmb{\mid}}}}\kern-.45em{\mathrm C}}}}}

\newcommand{\be}{\begin{equation}}
\newcommand{\bea}{\begin{eqnarray}}

\newcommand{\ee}{\end{equation}}
\newcommand{\eea}{\end{eqnarray}}

\newcommand{\non}{\nonumber \\}
\newcommand{\CR}{\non\cr}
\newcommand{\pa}{\partial}

\def\pa{\partial}

\newcommand{\aR}{{\alpha_R}}
\newcommand{\aI}{{\alpha_I}}
\newcommand{\bR}{{\beta_R}}
\newcommand{\bI}{{\beta_I}}

\begin{document}

\renewcommand{\thefootnote}{\fnsymbol{footnote}}

\makeatletter
\@addtoreset{equation}{section}
\makeatother
\renewcommand{\theequation}{\thesection.\arabic{equation}}

\rightline{}
\rightline{}



\begin{center}
{\LARGE \bf{\sc Nonabelian fluids and helicities}} 
\end{center} 
 \vspace{1truecm}
\thispagestyle{empty} \centerline{
{\large \bf {\sc Horatiu Nastase${}^{a}$}}\footnote{E-mail address: \Comment{\href{mailto:horatiu.nastase@unesp.br}}
{\tt horatiu.nastase@unesp.br}}
{\bf{\sc and}}
{\large \bf {\sc Jacob Sonnenschein${}^{b}$}}\footnote{E-mail address: \Comment{\href{mailto:cobi@tauex.tau.ac.il}}{\tt cobi@tauex.tau.ac.il}}
                                                        }

\vspace{.5cm}


\centerline{{\it ${}^a$Instituto de F\'{i}sica Te\'{o}rica, UNESP-Universidade Estadual Paulista}} 
\centerline{{\it R. Dr. Bento T. Ferraz 271, Bl. II, Sao Paulo 01140-070, SP, Brazil}}
\vspace{.3cm}
\centerline{{\it ${}^b$School of Physics and Astronomy,}}
\centerline{{\it The Raymond and Beverly Sackler Faculty of Exact Sciences, }} 
\centerline{{\it Tel Aviv University, Ramat Aviv 69978, Israel}}
 
\vspace{1truecm}

\thispagestyle{empty}

\centerline{\sc Abstract}

\vspace{.4truecm}

\begin{center}
\begin{minipage}[c]{380pt}
{\noindent 

In analogy with the non-Abelian gauge helicities  conserved in time for 
``null fields", that  we have defined previously,
in this paper we first define non-Abelian fluid helicities and then 
total non-Abelian helicities for combined non-Abelian fluid and gauge fields.  
For a U(N) group the helicities considered are 
for both gluonic-type fluids, composed of particles in the adjoint representation, 
and quark-type fluids, in the 
$N$-dimensional Cartan subalgebra.
We write down various Lagrangian formulations for the uncoupled and coupled systems. 
In each case we determine the equations of motion, 
symmetries, and a Hamiltonian formulation.
 Taking the velocity of the fluid in the adjoint representation, 
 we find that in the case of the gluonic fluid, we can write a 
non-Abelian gluonic fluid Euler-Yang-Mills equation that conserves the defined helicities, 
and comes from a Lagrangian. 

%

}
\end{minipage}
\end{center}

\vspace{.5cm}

\setcounter{page}{0}
\setcounter{tocdepth}{2}

\newpage

\tableofcontents
\renewcommand{\thefootnote}{\arabic{footnote}}
\setcounter{footnote}{0}

\linespread{1.1}
\parskip 4pt




\section{Introduction}
Non-Abelian fluids exist in nature. It is well known that heavy ion collisions, like those done in RHIC and the 
LHC, create a strongly-coupled quark-gluon plasma which is  a fluid. In fact it is very plausible that the very 
early universe before the deconfining/confining transition  has a form of a ball of non-Abelian fluid. 

The dynamics of non-relativistic perfect fluids is described by the Euler and continuity equations. 
One can couple them to external 
electromagnetic fields, as in \cite{Abanov:2021hio}.
``Null gauge fields'' admit  topologicaly non-trivial knot solutions associated with the  conservation of 
four helicities  ${\cal H}_{mm},{\cal H}_{me},{\cal H}_{em},{\cal H}_{ee} $(see \cite{ranada1990knotted}, 
\cite{Hoyos:2015bxa} and references therein). In \cite{Alves:2017ggb}, a map between these 
electromagnetic helicities  and 
fluid helicities ${\cal H}_f$ was derived. Correspondingly Hopfion fluid solution and other knot solutions  were 
written down.
 For the system of a fluid coupled to an external eletromagnetic fields, instead of the individual fluid 
helicity ${\cal H}_f$, 
the spatial CS form of the velocity, and the electromagnetic ${\cal H}_{mm}$ helicity, a total fluid-electromagnetic 
helicity ${\cal H}_{\rm tot}$ is conserved\cite{Nastase:2022aps} . 

In \cite{Nastase:2024scr}, we have defined helicities for Yang-Mills fields, generalizing the ones 
for electromagnetism. 
The notion of non-Abelian fluids was considered previously from various points of view, 
for instance in \cite{Bistrovic:2002jx}
and \cite{Cheung:2020djz}. 

But in this paper, we propose to think about it from the point of view of extending the notion of 
non-Abelian helicities from 
Yang-Mills to total ones for a combined system of non-Abelian fluid coupled to Yang-Mills fields. 

We  consider  fluids that are either "gluonic" in type, i.e., have an index $a$ in the adjoint, 
thought of as coming from a 
strongly-coupled gluon plasma, or "quark" in type, i.e., have an index $f$ 
that takes $N$ values for $U(N)$, specifically in the 
$N$-dimensional Cartan sub-algebra. 

We will first define the helicities, and then construct Euler equations, 
Lagrangians and classical solutions that conserve the 
helicities. 

In order to gain some insight, we review the non-relativistic, then the relativistic, Abelian fluids, from 
the point of view of a Lagrangian (with the corresponding symmetries) and Hamiltonian formulations. 
An essential point of the construction will be a Clebsch parametrization of the velocity, 
$u_\mu\propto a_\mu=\d_\mu\theta+\a \d_\mu \b$. 
A natural generalization would be to write a non-Abelian generalization in terms of $g,h\in {\cal G}$, 
and $u^\mu\propto a_\mu =g^{-1}\d_\mu g+\a h^{-1}\d_\mu h$, but we will find that it does not lead to a 
reasonable Euler equation (one that is covariant, and written in terms of $u^\mu$, not the Clebsch parameters), 
and so does not conserve our helicities.

Simple generalizations to the non-Abelian case of these Lagrangians are possible, and we find two of 
them, but one will give terms that depend only on the Clebsch parameters, and not on the full velocities, while
the other gives only one equation (a "traced" equation, instead of a full matrix equation, i.e., $N^2$ 
components, as we want). 

To continue, we will review the construction of \cite{Bistrovic:2002jx}, which does lead to an Euler-like equation
coupled to Yang-Mills fields through a Lorentz-like force, just that a single equation, not $N^2$ equations, 
one for each $a$ index.
The main difference between the approach of \cite{Bistrovic:2002jx} and our approach is  that we take  the 
fluid velocity in the adjoint representation, namely $u_\mu^a$. In both formulations the current is non-Abelian 
but in \cite{Bistrovic:2002jx} it is assumed that only the density carries an $a$ index and not the velocity, whereas 
in our formulation both $\rho$ and $u$ are in the adjoint representation.\footnote{Note that 
in \cite{Torrieri:2018qyx,Torrieri:2023udd} some problems were claimed to be identified with this construction,
specifically with a non-Abelian chemical potential, but with normal fluid velocity $u_\mu$.}

We then construct a Euler equation for non-Abelian gluonic fluids coupled to Yang-Mills fields, that does 
conserve our constructed helicities, and then find a Lagrangian formulation for it. We will also consider 
solutions for it. Finally, we present an attempt at an Euler equation for a non-Abelian "quark"-like (fundamental)
fluid coupled to Yang-Mills fields, but the result is somewhat trivial. 

Fluid dynamics has a long history of studies\footnote{A standard physics text on the subject is 
\cite{Landau:1986aog}.
A mathematical treatment is \cite{Arnold:2022}. The relation between Langrange and
Euler descriptions of a fluid is discussed by \cite{Salmon:1988};
see also \cite{Antoniou:2001pu}.}.
The topic of non-Abelian fluids was analyzed in the past by many authors. A Hamiltonian description appears in 
\cite{Morrison:1998zz},  non-Abelian fluids were studied in  \cite{Bistrovic:2002jx}, multi-fluids were discussed in 
\cite{Carter:2003im}, extensions to higher dimensions, internal symmetries and supersymmetry were explored in  
\cite{Jackiw:2004nm}, various aspects of anomalies in fluid dynamics were studied in \cite{Wiegmann:2024sqt}, 
\cite{Wiegmann:2024nnh}and \cite{Wiegmann:2024pxr}. 

The paper is organized as follows. In section 2 we review the Abelian helicities, and in section 3 we present 
our proposed non-Abelian helicities. In section 4 we review the Lagrangian formulations for the Abelian, 
non-relativistic and relativistic fluid dynamics, coupled to electromagnetism. 
In section 5 we present the non-Abelian fluid constructions. We start with the 
non-Abelian Clebsch parametrization
formulations, then review the formulation of  \cite{Bistrovic:2002jx}, after which we present the gluonic Euler 
equation coupled to Yang-Mills that conserves the corresponding helicities, with classical solutions, and 
end with the attempt at a "quark" (fundamental) fluid coupled to Yang-Mills.


\section{Abelian helicities}

In Maxwell's electromagnetism there is a sector of ''null fields''  $\vec E\cdot \vec B=0,$ $E^2=B^2$, for which on 
top of the ordinary $SO(4,2)$ symmetry, there are four types of conserved helicities that play the role of 
topological charges for knot solutions. In analogy to that, 
one can also define a null fluid with $(\vec v)^2=1$, for which the fluid dynamics admits a nontrivial conserved 
helicity\cite{Alves:2017ggb,Alves:2017zjt}
(in the absence of the null condition, the fluid conserves the fluid helicity only if it has boundary conditions
at infinity for fields vanishing fast enough). 
It turns out that a generalization of such a helicity is also conserved in the theory of a abelian charged fluid coupled 
to background electric and magnetic fields. In the following subsections we review these three types of helicities.

\subsection{Abelian gauge helicities}

In electromagnetism in vacuum, defining the vector potential $\vec{A}$ and its dual $\vec{C}$,
\be
\vec{E}=\vec{\nabla}\times \vec{C};\;\;\; \vec{B}=\vec{\nabla}\times \vec{A}\;,
\ee
one can define helicities as spatial integrals of spatial Chern-Simons (electric-electric and magnetic-magnetic)
or BF forms (mixed electric-magnetic) made up from $\vec{A}$ and $\vec{C}$, 
\bea
{\cal H}_{ee}&=&\int d^3x \vec{C}\cdot \vec{E}=\int d^3x \vec{C}\cdot \vec{\nabla}\times \vec{C}=\int d^3x \epsilon^{ijk}C_i\d_j C_k\;,\cr
{\cal H}_{mm}&=&\int d^3x \vec{A}\cdot \vec{B}=\int d^3x\epsilon^{ijk}A_i\d_j A_k\;,\cr
{\cal H}_{em}&=&\int d^3x \vec{C}\cdot\vec{B}=\int d^3x \epsilon^{ijk}C_i\d_j A_k\;,\cr
{\cal H}_{me}&=&\int d^3x \vec{A}\cdot\vec{E}=\int d^3x \epsilon^{ijk}A_i\d_k C_k\;,
\eea
which are conserved, by use of the Maxwell equations, if 
$\vec{E}\cdot \vec{B}\propto \epsilon^{\mu\nu\rho\sigma}F_{\mu\nu}
F_{\rho\sigma}=0$ and $\vec{E}^2-\vec{B}^2\propto F_{\mu\nu}F^{\mu\nu}=0$ (for "null" configurations, with 
null Riemann-Silberstein vector $\vec{F}=\vec{E}+i\vec{B}$, $\vec{F}^2=0$), since 
(see for instance \cite{Alves:2017ggb,Alves:2017zjt})
\bea
\d_t {\cal H}_{mm}&=&\int d^3x (\d_t \vec{A}\cdot \vec{B}+\vec{A}\cdot \d_t\vec{B})
=-\int d^3x(\vec{E}\cdot\vec{B}+\vec{A}\cdot(\vec{\nabla}\times\vec{E}))\cr
&=&-2\int d^3x \vec{E}\cdot\vec{B}=-\int d^3x \epsilon^{ijk}\d_i(E_jA_k)\;,\cr
\d_t {\cal H}_{ee}&=&\int d^3x (\d_t \vec{C}\cdot \vec{E}+\vec{C}\cdot \d_t\vec{E})
=-\int d^3x(\vec{B}\cdot \vec{E}+\vec{C}\cdot (\vec{\nabla}\times \vec{B}))\cr
&=&-2\int d^3x \vec{E}\cdot \vec{B}=-\int d^3x\epsilon^{ijk}\d_i(E_jA_k)\;,\cr
\d_t {\cal H}_{me}&=& \int d^3x(\d_t \vec{A}\cdot \vec{E}+\vec{A}\cdot\d_t\vec{E})
=\int d^3x(-\vec{E}\cdot\vec{E}+\vec{A}\cdot(\vec{\nabla}\times\vec{B}))\cr
&=&-\int d^3x(\vec{E}^2-\vec{B}^2)\;,\cr
\d_t {\cal H}_{em}&=& \int d^3x(\d_t \vec{C}\cdot \vec{B}+\vec{C}\cdot \d_t \vec{B})
=\int d^3x(-\vec{B}\cdot \vec{B}+\vec{C}\cdot (\vec{\nabla}\times \vec{E}))\cr
&=&\int d^3x (\vec{E^2}-\vec{B}^2)\;,
\eea
although, if one considers $\vec{A},\vec{C}\rightarrow 0$ sufficiently fast at the boundary, such that the boundary 
terms above are zero in the absence of the null conditions, 
${\cal H}_{ee}$ and ${\cal H}_{mm}$ are still conserved.

\subsection{ Fluid helicity}

In fluid dynamics, one can define (as defined and later studied by Moffat 
\cite{moffatt1969degree,moffatt1992helicity,moffatt1992helicity2})
the analog of $H_{mm}$, the fluid helicity defined as the spatial Chern-Simons form 
of the velocity, i.e., 
\be\label{Helicityf}
{\cal H}_f=\frac{1}{\Gamma^2}\int d^3x \vec{v}\cdot\vec{\nabla}\times \vec{v}
=\frac{m^2}{h^2}\int d^3x \vec{v}\cdot \vec{\omega}\;,
\ee
where $\Gamma$ is a normalization constant, here taken to be $=h/m$ 
in order for ${\cal H}$ to be integer valued
(in superconductivity this arises from the conditions on the wave function, but it was recently shown that this 
can be proved in generality, for any fluid \cite{Wiegmann:2024sqt}). 

Using the Euler equation, one finds that the fluid helicity is conserved 
(time independent) if the velocity goes to zero 
sufficiently fast at infinity (so that a boundary term involving the velocity vanishes).

\subsection{ The helicity of a charged fluid coupled to electromagnetic background}

For the Euler equation coupled to external electromagnetism, Abanov and 
Wiegmann \cite{Abanov:2021hio} considered a {\em total } 
fluid + electromagnetism helicity, defined in terms of the canonical momentum 
\be
\vec{\pi}=m\vec{v}+\vec{A}\;,
\ee
as the spatial Chern-Simons form of $\vec{\pi}$, i.e., 
\bea\label{helicitytot}
{\cal H}_{\rm tot}&=&\frac{1}{h^2}\int d^3 x\vec{\pi}\cdot \vec{\nabla}\times \vec{\pi}
=\frac{1}{h^2}\int d^3x \left[m^2\vec{v}\cdot \vec{\omega}
+\vec{A}\cdot \vec{B}+2m\vec{v}\cdot \vec{B}\right]\cr
&=&{\cal H}_f+{\cal H}_{mm}+2{\cal H}_{fm}\;,
\eea
where by partial integration $\int d^3x \vec{v}\cdot\vec{B}=\int d^3x \vec{A}\cdot\vec{\omega}$, 
so the 2 cross-terms 
(or "cross-helicities") are equal, giving what we called the cross fluid-electromagnetic helicity, ${\cal H}_{fm}$.

Abanov and Wiegmann \cite{Abanov:2021hio} (see also \cite{Nastase:2022aps} where solutions 
of the Euler coupled to 
electromagnetism with this total helicity were found) found that this total helicity is now 
conserved (time independent), 
\be
\frac{d}{dt}{\cal H}_{\rm tot}=0\;,
\ee
instead of the individual terms, a fact based on the Euler equation and 
the vanishing of $v_i, A_i$ at infinity. 

The proof was simple in a 4-dimensional formalism (even though the fluid 
theory was still non-relativistic) where one defined 
also a $\pi_0$ by 
\be
\pi_0=\Phi+A_0\;,\;\;\;-\Phi=\mu+\frac{m\vec{v}^2}{2}\;,
\ee
and with the usual 4-current $j^\mu=(\rho,\rho v^i)$, the Euler equation became
\be
j^\mu \Omega_{\mu\nu}=0\;,\;\; \Omega_{\mu\nu}\equiv \d_\mu \pi_\nu-\d_\nu\pi_\mu\;,
\ee
and the total helicity density 3-form was 
\be
h=\pi \wedge d\pi=\pi \wedge \Omega\Rightarrow dh=\Omega\wedge \Omega=0.
\ee

\section{ Non-abelian helicities}

A natural question is whether one can generalize the abelian helicities also to non-abelian helicities. 
In a similar manner to the abelian case, we will analyze first the issue for gauge fields and then for 
non-abelian fluids of different types.

\subsection { Non-abelian gauge helicities}

The electromagnetic helicities were generalized to non-Abelian (Yang-Mills) fields in 
\cite{Nastase:2024scr} as follows.

The non-Abelian singlet helicity extending ${\cal H}_{mm}$ was defined as the spatial Chern-Simons form of the 
non-Abelian field $A_i^a$, in the $A_0^a=0$ gauge, 
\bea
{\cal H}^{NA}_{mm}&=& \int d^3 x \Tr\left[ A\wedge dA+ \frac23 A\wedge A \wedge A\right]
=\int d^3x \Tr\left[A\wedge F 
-\frac{1}{3}A\wedge A\wedge A\right]\cr
&=&\int d^3 x \left [ A^a_i B^a_i 
-\frac13 \epsilon^{ijk}f^{abc}A^a_iA^b_jA^c_k\right ].\label{helNA}
\eea
which is conserved by the use of the YM equations, integration by parts and the $A_0^a=0$ gauge, 
\be
\d_t {\cal H}^{NA}_{mm} =2 \int d^3 x \vec E^a\cdot \vec B^a\;,
\ee
provided 
\be
\vec E^a\cdot \vec B^a=0\;.\label{EaBa}
\ee

The dual ${\cal H}_{ee}$ singlet helicity, the CS form of the dual $C_i^a$ field, 
\bea 
E^a_i&=& F^a_{i0}
= \frac12 {\epsilon_{i}}^{jk}(\pa_j C^a_k
-\pa_k C^a_j + {f^a}_{bc} C^b_j C^c_k)\cr
B^a_i&=& \frac12 \epsilon_{ijk} F^{ajk}
=-\d_t C_i^a\;,
\eea
is defined as
\bea
{\cal H}^{NA}_{ee}&=& \int d^3 x \Tr\left[ C\wedge dC+ \frac23 C\wedge C \wedge C\right]=\int d^3x \Tr\left[C\wedge \tilde F 
-\frac{1}{3}C\wedge C\wedge C\right]\cr
&=&\int d^3 x \left [ C^a_i E^a_i 
-\frac13 \epsilon^{ijk}f^{abc}C^a_iC^b_jC^c_k\right]\;, \label{heldualNA}
\eea
and is also conserved provided (\ref{EaBa}) is satisfied. 

On the other hand, the mixed singlet helicities, defined as the BF terms, 
\bea
{\cal H}^{NA}_{em}&=& \int d^3 x \Tr\left[C\wedge dA+ \frac12 A\wedge A \wedge C\right]
=\int d^3x \Tr\left[C\wedge F 
-\frac{1}{2}A\wedge A\wedge C\right]\cr
&=&\int d^3 x \left [C^a_i B^a_i 
-\frac12 \epsilon^{ijk}f^{abc}A^a_iA^b_jC^c_k\right]\cr
{\cal H}^{NA}_{me}&=& \int d^3 x \Tr\left[A\wedge dC+ \frac12 C\wedge C \wedge A\right]
=\int d^3x \Tr\left[A\wedge \tilde F 
-\frac{1}{2}C\wedge C\wedge A\right]\cr
&=&\int d^3 x \left [A^a_i E^a_i 
-\frac12 \epsilon^{ijk}f^{abc}C^a_iC^b_jA^c_k\right]
\eea
are not conserved,
\be
\d_t {\cal H}_{em}^{NA}=\int d^3x \left[\vec{E}^a\cdot \vec{E}^a-\vec{B}^a\cdot\vec{B}^a
+\frac{1}{2}\epsilon^{ijk}f_{abc}
(A_i^aA_j^b B_k^c-C_i^aC_j^bE_k^c)\right]\;.
\ee

Non-singlet non-Abelian helicities are defined similarly. The non-Abelian version of ${\cal H}_{mm}$ is 
again defined as the spatial Chern-Simons term of $A_i^a$, with $T^a$ inserted and in the $A_0^a=0$ gauge,
\bea
{{\cal H}^{NA}}_{mm}^a &=& \int d^3 x \Tr\left[T^a\left(A\wedge dA+ \frac23 A\wedge A \wedge A\right)\right]\cr
&=&\int d^3x \Tr\left[T^a\left(A\wedge F 
-\frac{1}{3}A\wedge A\wedge A\right)\right]\cr
&=&\int d^3 x\; {d^a}_{bc}\left [A^b_i B^c_i 
-\frac13 \epsilon^{ijk}f^{bde}A^c_iA^d_jA^e_k\right]\;,\label{helNA}
\eea
now conserved if ${d^a}_{bc}\vec{E}^b\cdot \vec{B}^c=0$. The same condition gives the  
conservation of the dual ${\cal H}_{ee}$ helicity,
\bea
{{\cal H}^{NA}}_{ee}^a &=& \int d^3 x \Tr\left[T^a\left(C\wedge dC+ \frac23 C\wedge C \wedge C\right)\right]\cr
&=&\int d^3x \Tr\left[T^a\left(C\wedge \tilde F 
-\frac{1}{3}C\wedge C\wedge C\right)\right]\cr
&=&\int d^3 x \;{d^a}_{bc}\left [C^b_i E^c_i 
-\frac13 \epsilon^{ijk}f^{bde}C^c_iC^d_jC^e_k\right].\label{helNA}
\eea

The mixed non-singlet helicities
\bea
{{\cal H}^{NA}}^a_{em}&=& \int d^3 x \Tr\left[T^a\left(C\wedge dA+ \frac12 A\wedge A \wedge C\right)\right]\cr
&=&\int d^3x \Tr\left[T^a\left(C\wedge F 
-\frac{1}{2}A\wedge A\wedge C\right)\right]\cr
&=&\int d^3 x\; {d^a}_{bc}\left [C^b_i B^c_i 
-\frac12 \epsilon^{ijk}f^{bde}A^c_iA^d_jC^e_k\right]\cr
{{\cal H}^{NA}}^a_{me}&=& \int d^3 x \Tr\left[T^a\left(A\wedge dC+ \frac12 C\wedge C \wedge A\right)\right]\cr
&=&\int d^3x \Tr\left[T^a\left(A\wedge \tilde F 
-\frac{1}{2}C\wedge C\wedge A\right)\right]\cr
&=&\int d^3 x \;{d^a}_{bc}\left[A^b_i E^c_i 
-\frac12 \epsilon^{ijk}f^{bde}C^c_iC^d_jA^e_k\right]\;,
\eea
are not conserved, though the condition ${d^a}_{bc}\left(\vec{E}^b\cdot \vec{E}^c-\vec{B}^b\cdot \vec{B}^c
\right)=0$ comes close to conserving them (they would be conserved for solutions that are 
duality symmetric, exchanging $\vec{A}^a$ 
with $\vec{C}^a$).\footnote{This is related to the fact that $\vec{A}^a$ and the dual $\vec{C}^a$  cannot 
be {\em simultaneously} defined in YM theory.}

In the following, we will want to define a non-Abelian version of a fluid, 
and couple it to Yang-Mills fields, just like the regular 
fluid was coupled to electromagnetism.

\subsection{ Non-Abelian fluid helicities}

The first step in attempting to generalize the concept of non-Abelian helicities to fluid dynamics is to 
define what is a non-Abelian fluid. 
There are several different ways to do it. The starting point is the non-Abelian fluid current that generates 
the non-Abelian global symmetry in the case that there are no non-Abelian gauge fields, and the local 
symmetry for the system coupled to non-Abelian gauge fields. Thus, the Abelian currents and their 
conservation law are uplifted  as follows. 
\be
j_\mu\rightarrow j^a_\mu \qquad \pa^\mu j_\mu^a= 0\ \ (global)\;\;\;{\rm or}\qquad D^\mu j_\mu^a \equiv 
\pa^\mu j_\mu^a+ f^{abc} {A^\mu}_b {j_\mu}_c=0 \ \ (local).
\ee

The Abelian fluid  current is built from the density $\rho$ and velocity  
\be
j^\mu = j  u^\mu \qquad j=\sqrt{j^\mu j_\mu}=\rho\sqrt{1-v^2}\;,
\ee
 so that for the non-relativistic case
\be
j^\mu = \rho( 1,\vec v).
\ee

For the non-Abelian current there are several options of how to build the non-relativistic  current. Possibilities 
one can consider are 
\be
(1)\  j_\mu ^a = \rho^a(1,\vec v)\;,\qquad  (2)\  j_\mu ^a = \rho^a(1,\vec v^a) \;,\qquad (3)\ j_\mu^a=j_\mu Q^a=
\rho (1,\vec{v})Q^a. 
\ee

Here a natural option would be for $a$ to be an adjoint index, but we will explore later also the possibility 
to have a fundamental index. 

In option (1) the flow velocity is the same for all the different non-Abelian  components of the density $\rho^a$. 
This would be a natural one if one is interested in a fluid where all the components move at the same 
velocity, except in \cite{Bistrovic:2002jx}, the variant (3) was considered, as we will review later.
In option (2) each component has different  velocity. There is no summation over the indices $a$ in (2). 

In the relativistic case the three options of the non-Abelian currrents can be written as
\be
(1)\ j_\mu^a= j^a u_\mu\;, \qquad (2)\ j_\mu^a= j^a u^a_\mu\;, \qquad (3)\ j_\mu^a =j_\mu Q^a\;,
\ee
where 
\be
j^a= \sqrt{j^a_\mu {j^\mu}^a} =\rho^a\sqrt{1-v^2}\;,
\ee	
namely, using the singlet velocity and with no summation over $a$.


Thus, to get  fluid helicities which are the analogs of the singlet and non-singlet 
non-Abelian helicities we need to adopt option (2). 

The singlet helicity reads
\bea
{\cal H}^{NAs}_{\rm fl}&=& \int d^3 x \Tr\left[ {\bf V}\wedge d{\bf V}+ 
\frac23 {\bf V}\wedge {\bf V} \wedge {\bf V}\right]\cr
&=& \int d^3 x \left [ {V}^a_i \epsilon^{ijk}\pa_j{V}_k^{a} 
-\frac13 \epsilon^{ijk}f^{abc}{V}^a_i{V}^b_j{V}^c_k\right ].
\eea
Here ${\bf V}=V_i^aT_adx^i $ is a non-Abelian spatial one-form, and ${\bf W}=d{\bf V}$.

In a similar manner to the non-Abelian adjoint helicities for the gauge fields, also for the fluid  
case one has to insert a matrix $T^a$  into the trace as follows. 
\bea
{{\cal H}^{NAa}}_{fl}^a &=& \int d^3 x \Tr\left[T^a\left({\bf V} \wedge d{\bf V}
+ \frac23 {\bf V}\wedge {\bf V} \wedge {\bf V}\right)\right]\cr
&=&\int d^3x \Tr\left[T^a\left(V\wedge {\bf W} 
-\frac{1}{3}{\bf V}\wedge {\bf V}\wedge {\bf V}\right)\right]\cr
&=&\int d^3 x\; {d^a}_{bc}\left [{V}^b_i { W}^c_i 
-\frac13 \epsilon^{ijk}f^{bde}{ V}^c_i{ V}^d_j{ V}^e_k\right]\;.\label{helNAf}
\eea

\subsection{Helicities of a ``Gluonic'' fluid coupled to gauge fields}

Next we consider the non-Abelian helicities of the fluid coupled to gauge fields. We will treat two separate cases.

It is not clear what a non-Abelian fluid can correspond to, but in this subsection we are considering 
the possibility of a fluid made up 
of only gluons, so a "gluonic fluid", interacting with the Yang-Mills fields themselves. 

In other words, in the case of a strongly coupled plasma like the one obtained in heavy 
ion collisions (at RHIC or Alice at the 
LHC), if we ignore the quarks, we should have a gluon plasma, and 
because of the strong coupling, this can be described 
as a fluid. And yet, there is still the Yang-Mills field around, and it 
should interact with the gluonic fluid, giving the model we are 
after. 

There are many models one can have about the non-Abelian fluid. 
Instead of starting with a Lagrangian or an equation of 
motion, we find it more convenient to start with a definition of helicity, and look for models that conserve it in time. 

If $a$ is an index in the adjoint representation, define then the natural 
generalization of the Abelian case, namely option (2) for the non-coupled fluids, i.e.,
a velocity $\vec{v}_a$, fluid density $\rho_a$, such that 
\be
\vec{j}_a=\rho_a\vec{v}_a\;,
\ee
and mass $m_a$ of the fluid particle, chemical potential $\mu_a$ of the same, and charge $Q_a$, so 
\be
\Pi_\mu=\Pi^a_\mu T_a\;,\;\;
\vec{\Pi}_a=m_a\vec{v}_a+Q_a\vec{A}_a\;,\;\;
\Pi_{0a}=-\mu_a-m_a\frac{\vec{v}^2_a}{2}+A_{0a}\;,
\ee
where there is no sum over $a$.
Note that 
\be
\rho_a(\vec{x},t)=\sum_n m_a \delta^3(\vec{x}-\vec{x}_{an}(t))\;,
\ee
where $\vec{x}_{an}(t)$ is defined by some initial position $\vec{x}_{an}$ and some velocity $\vec{v}_{an}(t)$,
so we can consider the case that all $m_a$ are equal, $m_a=m$, 
as long as the $\vec{v}_a$ are different for different $a$ 
(such that $\rho_a$ are all different). 
We can also consider all $Q_a$ to be equal, meaning all fluid 
particles interact the same way with the Yang-Mills fields, though
we will leave it as it is for now.

In terms of them, define the forms
\be
\Pi=\Pi^ a T_a\;,\;\; \Omega=\Omega^ aT_a\;,\;\; \Omega=d\Pi+\Pi \wedge \Pi\;, 
\ee
and thus the non-Abelian helicity densities
\be
h^a=\Tr\left[T^ a\left(\Pi \wedge d\Pi +\frac{2}{3}\Pi\wedge \Pi\wedge \Pi\right)\right]\;,
\ee
as well as the singlet non-Abelian helicity density $h$, with $T^a$ replaced by the identity, 
such that
\be
dh^ a=\Tr\left[T^ a\left(\Omega\wedge \Omega\right)\right]\;,
\ee
and similarly for $dh$.

The total singlet non-Abelian helicity is then 
\bea
{\cal H}^{NA}_{\rm total}&=& \int d^3 x \Tr\left[ \Pi\wedge d\Pi
+ \frac23 \Pi\wedge \Pi \wedge \Pi\right]=\int d^3x \Tr\left[\Pi\wedge 
\Omega-\frac{1}{3}\Pi\wedge \Pi\wedge \Pi\right]\cr
&=&\int d^3 x \left [ \Pi^a_i \epsilon^{ijk}\Omega^a_{jk} 
-\frac13 \epsilon^{ijk}f^{abc}\Pi^a_i\Pi^b_j\Pi^c_k\right ]\cr
&=&{\cal H}^{NA}_{mm}+{\cal H}^{NA}_{fm}+{\cal H}^{NA}_f\;,\label{helNAPi}
\eea
where ${\cal H}^{NA}_{mm}$ is the magnetic Yang-Mills helicity, ${\cal H}^{NA}_f$ is a non-Abelian analog of the 
fluid helicity, and ${\cal H}^{NA}_{fm}$ is a mixed term.
The total non-singlet non-Abelian helicity is 
\bea
{{\cal H}^{NA}}_{mm}^a &=& \int d^3 x \Tr\left[T^a\left(\Pi\wedge d\Pi
+ \frac23 \Pi\wedge \Pi \wedge \Pi\right)\right]\cr
&=&\int d^3x \Tr\left[T^a\left(\Pi\wedge \Omega 
-\frac{1}{3}\Pi\wedge \Pi\wedge \Pi\right)\right]\cr
&=&\int d^3 x\; {d^a}_{bc}\left [\Pi^b_i \Pi^c_i 
-\frac13 \epsilon^{ijk}f^{bde}\Pi^c_i\Pi^d_j\Pi^e_k\right]\cr
&=&{{\cal H}^{NA}}_{mm}^a+{{\cal H}^{NA}}_{fm}^a+{{\cal H}^{NA}}_f^a\;,\label{helNAPi}
\eea
with a similar definition for the 3 terms.

We will find that these helicities are conserved by a non-Abelian generalization of the Euler equation.

\subsection{Fundamental ("quark fluid") helicities}

One could think perhaps of a "quark fluid" instead of the "gluon fluid", i.e., 
that only the quarks interact strongly and form a 
fluid, and then this fluis interacts with the resulting Yang-Mills field of the gluons. 

Since the quarks $q$ are in the fundamental representation, 
they should have an index $i$ in the fundamental representation. 
But, moreover, they should have another index, $f$, that labels the 
type of fluid particle, and the particle is associated with a 
$T_f$ in the Cartan subalgebra of the Lie algebra. 
This is a situation similar to the one considered by \cite{Bistrovic:2002jx}, 
though the rest of the construction is different. 

We consider then the canonical momentum (we have specialized to $m_f=m$, 
but by a similar argument as in the 
gluon case from the previous subsection, this doesn't change anything)
\be
\vec{\Pi}_f=m\vec{v}_f\sum_a \delta^{fa}T_a+\sum_{\tilde b}Q_{f\tilde b}\vec{A}^{\tilde b} 
T_{\tilde b}\equiv m\vec{v}_f+\vec{A}_f\;,
\;\;\Pi_f=\vec{\Pi}_fd\vec{x}\;,
\ee
where 
\be
Q_{f\tilde b}\equiv\sum_{i,j}\bar q_{fi}(T_{\tilde b})_{ij}q_{fj}
\ee
is the charge coupling of a $\bar q q$ pair to Yang-Mills gluons. 

$T_f$ must be in the Cartan subalgebra (commuting elements), which is $N$-dimensional for $U(N)$, and then 
$T_{\tilde b}$ is taken to be another subset of $T_a$.

We can define then the total helicity 
\bea
{\cal H}'_{\rm tot}&=&\sum_f\int d^3x \Tr\left[\Pi_f\wedge d \Pi_f +\frac{2}{3}\Pi_f\wedge \Pi_f\wedge \Pi_f \right]\cr
&=&m^2 \int d^3x \sum_f\vec{v}_f\cdot \vec{\omega}_f+
\int d^3x\sum_f\Tr\left[A_f\wedge d A_f +\frac{2}{3} A_f\wedge A_f \wedge A_f\right]+{\rm more}.\cr
&&\label{htot}
\eea

In order to not obtain extra terms (the "more" above) we would need to have the conditions
\be
\Tr[T_f\tilde T_{\tilde a}]=0\;,\;\; \
\Tr[T_{f_1}[T_{f_2},T_{f_3}]]=0\;,\;\;
\Tr[T_{f_1}[\tilde T_{\tilde a},\tilde T_{\tilde b}]]=0\;,
\Tr[[T_{f_1},T_{f_2}]\tilde T_{\tilde a}]=0.\label{comm}
\ee

If we choose $T_{\tilde a}$ to be in the Lie algebra, except the Cartan subalgebra, we only find the extra term
\be
\int d^3x 2m\sum_f\Tr[v_f\wedge A_f\wedge A_f].
\ee

Another possibility is that we define the sum over $f$ already in $\Pi$, so 
\be
\Pi=\sum_f \Pi_f\;,\;\;\;
A\equiv \sum_f Q_{f\tilde b} A^{\tilde b}T_{\tilde b}\;,
\ee
in which case we would define the total helicity as 
\bea
{\cal H}''_{\rm tot}&=&\int d^3x \Tr\left[\Pi\wedge d \Pi +\frac{2}{3}\Pi\wedge \Pi\wedge \Pi \right]\cr
&=&m^2 \int d^3x \sum_f\vec{v}_f\cdot \vec{\omega}_f+
\int d^3x\sum_f\Tr\left[A\wedge d A +\frac{2}{3} A\wedge A \wedge A\right]\cr
&&+2m\int d^3x \Tr[v\wedge A \wedge A].
\eea

But the question is: is either ${\cal H}'_{\rm tot}$ or ${\cal H}''_{\rm tot}$ 
conserved by a non-Abelian Euler equation with an 
index $f$? We will see that, unfortunately, while we can conserve ${\cal H}'_{\rm tot}$, 
the answer is somewhat trivial, 
since the fluid types $f$ are non-interacting.

\section{Lagrangian  formalism of Abelian fluid dynamics}

At this point we would like to place the conserved helicities  in the framework of Lagrangian formulations of fluid 
dynamics. We begin with the ordinary uncharged fluid first in the non-relativistic limit 
and then in relativistic regime. 
We then present a Lagrangian formulation for a charged fluid coupled to an Abelian gauge field again both in the 
nonrelativistic and relativistic domains.

\subsection{The nonrelativistic fluid}

First, let us review the Lagrangian formulation and the corresponding  equations of motion of  the nonrelativistic 
fluid that were considered in   \cite{Bistrovic:2002jx}. 
We then address the issues of the dimensions of the fields, the  
Hamiltonian formulation, and the symmetries. 
The equations of motion and the conservation laws are shown to be compatible with the continuity  and Euler 
equations of the fluid.

\subsubsection{ The Lagrangian density}
 
 The  nonrelativistic fluid degrees of freedom 
were  characterized  in \cite{Bistrovic:2002jx}  
by the fluid density $\rho$, the flow velocity $\vec v$ and the auxiliary gauge field $a_\mu$.
The system was  defined  by the following  Lagrangian density 
\be
{\cal L}(\rho,\vec v, a_\mu)=-j^\mu a_\mu +\frac{1}{2}\rho \vec{v}^2-V(\rho)\;,\label{Lagdennr}
\ee
where $V$ is some potential giving a pressure (or force), $j^\mu$ is the 4-current 
defined as  
\be
j^\mu=(c\rho,\rho\vec{v})\;.\;\;\;\label{current}
\ee

The variation with respect to $a_\mu$ leads to $j_\mu=0$ clearly not describing a fluid.
 However, if one parametrizes the gauge field \cite{Bistrovic:2002jx} in the Clebsch parametrization 
\be
a_\mu=\d_\mu\theta+\a\d_\mu\b\;,
\ee
 so that now in fact the Lagrangian density is ${\cal L}( \rho, \vec v, \theta,\alpha,\beta)$
and it takes the explicit form
\be
{\cal L}( \rho, \vec v, \theta,\alpha,\beta)= -\rho( \pa_t\theta+ \a\pa_t\b)-\rho\vec v 
\cdot(\vec\nabla\theta +\a\vec\nabla\b) + \rho\frac{\vec{v}^2}{2}- V\;,
\ee
which can be rewritten as 
\be
{\cal L}=\rho\left[\frac{\vec{v}^2}{2}-\frac{V}{\rho}-\frac{D\theta}{Dt}-\a\frac{D\b}{Dt}\right]\;,
\ee
where 
\be
\frac{D\theta}{Dt}= \pa_t\theta +\vec v\cdot \vec\nabla\theta \qquad \frac{D\b}{Dt}
= \pa_t\b +\vec v\cdot \vec\nabla\b. \label{Lagnr}
\ee

A variant, though  equivalent, action  was proposed in \cite{Nastase:2023rou}, 
\bea
\int d^4x {\cal L}&=&\int d^4x \left\{\frac{\rho\vec{v}^2}{2}-\rho\d_0\tilde\mu
+\phi\left(\frac{D\rho}{Dt}+\rho\vec{\nabla}\cdot\vec{v}\right)-\rho\a_a\frac{D\b^a}{Dt}\right\}\CR
&=&\int d^4x\left\{\rho\left[\frac{\vec{v}^2}{2}-\d_0\tilde\mu-\frac{D\phi}{Dt}-\a^a \frac{D\b^a}{Dt}\right]\right\}\;,
\eea
where the second form was obtained by partial integration, and it is then equivalent to 
(\ref{Lagnr}), with $-\d_0\tilde\mu$
being the same term as $-V/\rho$.


\subsubsection{ Dimensions of the fields}


One can
assign dimensions to the various fields 
as
\bea
&&[a_\mu]=1,\qquad [j_\mu]=3, \qquad [\rho]=3,\qquad [v_\mu]=0,\cr
&& \frac12\rho v^2\rightarrow \frac12 
m\rho v^2 \qquad V(\rho)\rightarrow m^4V\left(\frac{\rho}{m^3}\right) .
\eea

In this assignment $\rho$ has the meaning of a number density or charge density. 
If however we want that the dimension of $\rho$ to  reflect the fact that it is mass density namely 
$[\rho]=4$, and that as before $[j_\mu]=3$ and  $[v_\mu=0]$, we have to modify the relation between 
$j_\mu$ and $v_\mu$ as  $j_\mu=\frac{\rho}{m} v_\mu$.

\subsubsection{The equations of motion}

\begin{itemize}
\item
The equation of motion associated with the variation  of  $\vec{v}$ 
gives the velocity in terms of  the Clebsch parametrization, 
\be
\vec{v}=\vec{\nabla}\theta+\a\vec{\nabla}\b.\label{vClebsch}
\ee
\item
The variation with respect to $\theta$ yields the  equation of motion 
\be
\pa_\mu j^\mu=\d_t\rho+\vec{\nabla}\cdot( \rho\vec{v})=0\;,\label{vartheta}
\ee
which is  the continuity equation.
\item
The equations of motion that follow from the variation of  $\b,\a$ are 
\bea
j^\mu\d_\mu \a&=& \rho(\dot \a+\vec{v}\cdot\vec{\nabla}\a)\equiv \rho\frac{D\a}{Dt}=0\cr
j^\mu\d_\mu \b&=& \rho(\dot \b+\vec{v}\cdot\vec{\nabla}\b)\equiv \rho\frac{D\b}{Dt}=0\;,\label{varab}
\eea
where in the first equation we have used the continuity equation $\pa_\mu j^\mu=0$.
\item
Varying $\rho$ gives the Bernoulli-type equation, 
\be
\dot \theta+\a\dot\b+\frac{\vec{v}^2}{2}=-\frac{\delta}{\delta \rho}\int d^d\vec{r}
V\equiv -1-\frac{P}{\rho}\;,\label{Bernuli}
\ee
and the Euler equation then is obtained from a combination of these equations, namely from a derivative of the 
Bernoulli equation, using the $\a,\b$ equations.

\end{itemize}


\subsubsection{Hamiltonian formulation}

The  momentum conjugate to a given field $\phi_i$ is given by 
\be
\pi_i=\frac{\delta{\cal L}}{\delta \dot\phi_i}\;,
\ee
and the corresponding Hamiltonian density ${\cal H}$ is
\be
{\cal H}=\sum_i \pi_i\dot \phi_i-{\cal L}.
\ee

For the Lagrangian density (\ref{Lagdennr}) the only nontrivial conjugate momenta are
\be
\pi_\theta= -\rho,\qquad \pi_\b= -\a\rho,
\ee 
and all the other vanish,
\be
 \pi_\a=\pi_\rho=\pi_{\vec v}=0.
\ee

The associated Hamiltonian density is
\be
{\cal H} =-\rho\dot\theta - \rho\a\dot\b-{\cal L}=\rho\vec v \cdot(\vec\nabla\theta +\a\vec\nabla\b) 
+\frac12 \rho {\vec v}^2 + V(\rho).\label{Hamnr} 
\ee

This Hamiltonian density, not surprisingly,  coincides with $T_{00}$ that will be derived as a 
Noether current in (\ref{Tmunu}) and from the coupling to background metric, 
as we will do later, in (\ref{relagrav}).

From the Hamiltonian formulation one can proceed and canonically quantize the system.  The fact that 
$\pi_\rho=\pi_{\vec v}=\pi_\a=0$ implies that these are constraints and the 
canonical quantization has to proceed using the Dirac brackets. 
We will leave this to a future research project.\footnote{Although the system was only classically described, one 
could imagine, for instance, a quantum fluid, as in our motivation, the strongly-coupled quark-gluon plasma
(sQGP) obtained at RHIC and ALICE.}


\subsubsection{ Symmetries}

The symmetries of the action of the nonrelativstic fluid are:
\begin{itemize}
\item
Invariance under space-time translations. The corresponding Noether current, 
the energy momentum tensor, is given by
\be 
T_{\mu\nu}=-\sum_{\phi^q=\theta,\b,\rho,\vec{v},\a}\frac{\d {\cal L}}{\d \d^\mu \phi^q}\d_\nu \phi^q+{\eta}^{\mu\nu}
{\cal L}\;,
\ee
so
\bea
T_{00} &=& -\rho(\pa_t \theta +\a\pa_t\b)-{\cal L}= \frac12 \rho v^2 + V \CR
T_{0i} &=& -\rho(\pa_i\theta+\a\pa_i\b)= -\rho v_i\Rightarrow {T^0}_i=\rho v_i\CR
T_{i j}&=& +\rho(\pa_i\theta+\a\pa_i\b)v_j +\delta_{ij}{\cal L}\cr
&=& +\rho v_i v_j -\delta_{ij}\left(-\rho\frac{\delta}{\delta \rho}\int V 
+ V\right )\;,\label{Tmunu}
\eea
where in ${\cal L}$ we have used the equation of motion for $\vec{v}$, $\vec{v}=\vec{\nabla}\theta
+\a\vec{\nabla}\b$, and the Bernoulli equation. Note that the bracket multiplying $-\delta_{ij}$ becomes in the 
nonrelativistic limit $-P+...$.

Since in the nonrelativistic limit
\be
V\simeq \rho+...\;,
\ee
the conservation of the energy-momentum tensor reads
\bea
0=\pa_t T_{00}-\pa_i T_{i0} &=& \pa_t\left(\frac{\rho v^2}{2} + V\right)+\pa_i(\rho v_i) \cr
&\simeq&[\pa_t\rho +\pa_i(\rho v_i)] + \pa_t(\rho v^2)\CR
0=-\pa_t T_{0i}+\pa_j T_{ji} &=& \rho[\pa_t v_i+ v_j\pa_j v_i -f_i] + v_i[\pa_t\rho +\pa_j(\rho v_j)]\;,
\eea
the first being the continuity equation (plus higher order terms), and the second a linear combination of the 
continuity equation, and the Euler equation with force per particle $f_i=\d_i\left(V-\rho\frac{\delta}{\delta \rho}
\int V\right)$.
 
\item
Invariance under rotations. The   angular momentum charge is given by
\be
J_{ij}=\int d^3 x J_{0ij}=\int d^3 x( T_{0i}x_j-T_{0j} x_i)=\int d^3x\rho(v_i x_j- v_j x_i).
\ee
Needless to say that the nonrelativistic action is not invariant under boosts.
\item
The action is invariant under the  time reversal symmetry 
\be
t\rightarrow -t \qquad \theta(-t)=-\theta(t)\qquad \b(-t)\rightarrow -\b(t)\;,
\ee 
where obviously the flip of sign of $\b$ can be interchanged by a flit of signs of $\a$, these latter two 
implying $\vec{v}\rightarrow -\vec{v}$, as they should.
\item 
In a similar manner, the parity transformations 
\be
\vec x\rightarrow -\vec x \qquad \theta(-\vec x)=-\theta(\vec x)\qquad \b(-\vec x)\rightarrow -\b(\vec x)
\ee
also keep the action invariant.
\item
A shift symmetry of $\theta$,
\be
\theta\rightarrow \theta + a.
\ee
The associated conserved current is the fluid current, given in (\ref{current}),
\be
j_{(\theta)\mu} =\rho v_\mu= (\rho,\rho \vec v) \;,
\ee
whose conservation we saw that corresponds to the $\vec{v}$ equation of motion. 
\item
A shift symmetry of $\b$,
\be
\b\rightarrow \b + b.
\ee
The associated current is 
\be
j_{(\b)\mu}= \alpha\rho v_\mu=\a (\rho,\rho \vec v) .
\ee
The conservation of this current is the equation of motion
\be
\pa^\mu {j_\mu}_\b= \pa^\mu (\alpha\rho v_\mu)=\rho\frac{D\a}{Dt} =0.
\ee
\end{itemize}


\subsubsection{Nonrelativistic fluid coupled to electromagnetism}

The coupling to external electromagnetism was  obtained in \cite{Bistrovic:2002jx} by replacing 
$a_\mu\rightarrow a_\mu+A_\mu$, 
\be\label{atA}
a_\mu\rightarrow a_\mu+A_\mu =\d_\mu\theta+\a\d_\mu \b+A_\mu\;,
\ee
where $A_\mu$ is 
the electromagnetic field.

The physical meaning of this coupling is that the fluid current $j^\mu$ takes the role of the electromagnetic 
current $j^\mu_{EM}$, and in particular the fluid density $\rho$ is now the electric charge density.

The equations of motion of the coupled system follow again from the variations of 
$\rho, \vec v, \theta, \a$ and $\b$:
\begin{itemize} 
\item
A variation with respect to $A_\mu$ is relevant only when the gauge fields are dynamical, 
namely by adding the $-\frac14 F_{\mu}F^{\mu\nu}$ term to the Lagrangian density ${\cal L}$. 
In that case one gets the familiar Maxwell's equations $\pa^\mu F_{\mu\nu}= j_\nu$. Here we consider 
a coupling to an electromagnetic background with no backreaction, so we should not vary the Lagrangian 
density with respect to $A_\mu$.

\item
The variation with respect to $\theta,\a$ and $\b$ yield the same equations as for the uncoupled system.
\item
From the variation with respect to $\vec v$ one finds
\be
\vec{v}=\vec{\nabla}\theta+\a\vec{\nabla}\b +\vec{A}.\label{vClebschA}
\ee
 \item
The variation with respect to $\rho$ yields the  same equation as (\ref{Bernuli}).

Again if we take the $\pa^i$ derivative and use the other equations of motion we get
 the electromagnetic-coupled Euler equation (with $q=1$, $m=1$),
\bea
\d_tv^i+(\vec{v}\cdot\vec{\nabla})v^i&=&E^i+[\vec{v}\times \vec{B}]^i-\d^i\frac{\delta}{\delta\rho}\int dr V\cr
\frac{\delta}{\delta\rho}\int dr V&=&1+\frac{P}{\rho}.
\eea
\end{itemize}

The symmetries of the coupled system are the same as those of the uncoupled one. A realization of 
gauge invariance requires turning the gauge fields  into dynamical fields and incorporating the 
backreaction of the fluid on the electric and magnetic fields. 


\subsection { The relativistic fluid}

The  relativistic fluid  is by definition  described  by  a Lorentz invariant action. The Lagrangian density is 
a functional of $a^\mu= \pa^\mu\theta +\a\pa^\mu\b$. The fluid degrees of freedom  can be expressed 
either in terms of  $j^\mu$ or in terms of $\rho$ and $u^\mu$. These two options yield 
different equations of motion.   
It is convenient to define
\be
j\equiv \sqrt{j^\mu j_\mu/c^2}=\rho\sqrt{1-\vec{v}^2/c^2}= \frac{\rho}{\gamma}\;,\qquad j_\mu = j u_\mu.
\ee

Using this definition, \cite{Bistrovic:2002jx} proposed the  Lagrangian density  
\be\label{Lagrel}
{\cal L}=-j^\mu a_\mu -f(j)\;,\;\;\;
\ee
where $f$ is an arbitrary function of the relativistic number density $j$.  
For the free case, corresponding to the  nonrelativistic Lagrangian with $V=0$, 
$f$ takes the form of 
\be
f(j)=f_0(j)=jc^2\;; \qquad {\cal L}=-j^\mu a_\mu -jc^2.
\ee

   In the presence of a potential $V$, the non-relativistic limit of this action has
\be
f(j)=jc^2+V(j)\;,\;\;\; j\simeq \rho-\frac{1}{2c^2}\rho\vec{v}^2\;,\;\; u^\mu \simeq (c,\vec{v}).
\ee 
 
The dimension assignment in this relativistic case is:
\begin{itemize}
\item 
Assuming the dimension of the auxiliary gauge field is the canonical $[a_\mu]=1$, then again  $[j_\mu]=3$,
so that the $f(j)$ term has to be  of the form
\be
f(j)\rightarrow m^4 f\left(\frac{j}{m^3}\right).
\ee
\end{itemize}


\subsubsection{Equations of motion}

\begin{itemize}
\item
The variation with respect to $\theta$ is   the same as in (\ref{vartheta}). This yields, as before, 
the continuity equation. The variations with respect to  $\a$ and $\b$ are 
given in  (\ref{varab}). 
\item
At this point there are two options for what are the fields with respect to which one varies the 
Lagrangian density: (i) variation with respect to $j^\mu$; (ii) variation with respect to $\rho$ and $u^\mu$.
The former is manifestly relativistic invariant and the latter is not, since $\rho$ is a zero component 
of a vector and not a scalar.  
These two options are different, since  in the first option the components $j^i$ are composite of two fields, 
$\rho$ and $v^i$ of the second one, and a variation with respect to products of fields is different than 
the separate variations of them.
\item
The variation with respect to $j^\mu$ yields
\be
a_\mu =-\frac{u_\mu}{c^2}f'(j)=-\frac{j_\mu}{ j c^2}f'(j).\label{eqnjr}
\ee

It was shown in \cite{Bistrovic:2002jx} how to get from this equation of motion, the relativistic Euler equation.
One first takes the curl of the last equation
\be
\pa_\mu a_\nu - \pa_\nu a_\mu= \frac{1}{c^2}(\pa_\nu(u_\mu f'(j))-\pa_\mu(u_\nu f'(j))) .
\ee
The left hand side is equal to $\pa_\mu\a\pa_\nu \b- \pa_\nu\a\pa_\mu \b$, 
which vanishes when projected on $u^\mu$  due to (\ref{varab}). In this way we get the relativistic Euler equation
\be
u^\mu\pa_\mu\left[u_\nu f'(j)\right] -\pa_\nu f'(j)=0.\label{relEuler}
\ee
\item
Next we check the separate variation with respect to $\rho$ and $u^\mu$.
The variation of $\rho$ is
\be
-\frac{u^\mu a_\mu}{\gamma} -\frac{f'(j)}{\gamma}=0.
\ee

This equation is the same as the equation (\ref{eqnjr}), upon multiplying it with $u^\mu$.

\item
The variation with respect to $\vec v$ of the free action yields
\be
-j \vec a + \frac{\rho \vec v}{\sqrt{1-\frac{v^2}{c^2}}}=0 \rightarrow \qquad  \vec a
= \frac{ \vec v}{(1-\frac{v^2}{c^2})}.
 \ee
 
 In the non-relativistic limit this coincides with (\ref{vClebsch}).
\end{itemize} 


\subsubsection{Hamiltonian formulation}

As in the nonrelativistic case, here also only $\theta$ and $\beta$ have  nontrivial conjugate momenta,
$\pi_\theta=-\rho$ and $\pi_\beta=-\rho\a$,  so that the corresponding Hamiltonian density is 
\be
{\cal H} =-\rho\dot\theta - \rho\a\dot\b-{\cal L}=  \rho {\vec v}^2 +f(j) \;,
\ee
which indeed in the nonrelativistic limit goes into (\ref{Hamnr}).

\subsubsection{Symmetries}

\begin{itemize}

\item 
The symmetries for this system are the same as those of the non-relativistic case, 
apart from the fact that now the action is also invariant under boosts.

\item
The energy-momentum tensor that follows from the Noether procedure  reads
\bea
T_{\mu\nu}&=&-\sum_{\phi_q=\theta,\b}\frac{\d {\cal L}}{\d (\d^\mu\phi_q)}\d_\nu \phi_q+\eta_{\mu\nu}{\cal L}\cr
&=& -\frac{u_\mu u_\nu}{c^2} j f'(j) + \eta_{\mu\nu}( jf'(j)-f(j))\;,
\eea
where we have used (\ref{eqnjr}).
As was shown in \cite{Bistrovic:2002jx}, the conservation of the energy-momentum tensor  translates 
to the continuity and relativistic Euler equations as follows:
\be
\pa^\mu T_{\mu\nu}= -\frac{1}{c^2}\left [\pa^\mu(j u_\mu)u_\nu f'(j) + j\left\{ u_\mu \pa^\mu(u_\nu)f'(j)
-c^2 \pa_\nu f'(j)\right\}\right ].
\ee
It is easy to see that the vanishing of the first term corresponds to the continuity equation, and the 
vanishing of the second term is the relativistic Euler equation.
\end{itemize}


\subsection{The coupling of the relativistic fluid to electromagnetism}

The coupling of a charged relativistic fluid to electromagnetic fields is done in exactly the same 
way as for the nonrelativistic fluid,
namely introducing the replacement (\ref{atA}) into the action (\ref{Lagrel}), so
\be
{\cal L}=-j^\mu (a_\mu+A_\mu) -f(j)\;.
\ee

The procedures of writing the equations of motion, the Hamiltonian formulation, the symmetries 
and the conservation laws follow the same line as before.

The right-hand side of the Euler equation (\ref{relEuler}) becomes just $c^2j^\mu F_{\mu\nu}$, the 
correct Lorentz force coupling to electromagnetism.


\section{The Lagrangian formalism for non-Abelian fluid dynamics}

\subsection{Preliminary discussion}

Upon uplifting an ordinary fluid and a fluid that couples to Abelian gauge fields to non-Abelian fluid and a 
fluid that couples to  non-Abelian gauge fields, one has to devise an uplift to the basic fields of the fluid 
theory namely: $\rho$, the fluid density, $u^\mu$, the flow velocity, $j^\mu$, the fluid current and $\theta,\a$ 
and $\b$, the Clebsch parametrization of the auxiliary gauge fields $a_\mu$. In this section we propose 
two prescriptions of performing the uplift for non-Abelian fluids, first without coupling to non-Abelian 
gauge fields and then with such coupling. 

\begin{itemize}
\item
Obviously, the current and the auxiliary gauge field have to be mapped into elements of the algebra 
of the non-Abelian  group, namely
\be
j_\mu \rightarrow j_\mu=j_\mu ^a T^a \qquad  a_\mu \rightarrow a_\mu=a_\mu ^a T^a.
\ee
\item It is thus clear how to uplift $j$
\be
j\equiv 2\sqrt{\Tr[j^\mu j_\mu]}= \sqrt{{j^\mu}^a j^a_\mu}.
\ee
\item
The uplift of the auxiliary gauge fields $a_\mu$ implies a prescription of the  uplift to the Clebsh fields 
$\theta,\a$ and $\b$. The prescription should  be such that when one goes back from a non-Abelian group 
${\cal G}$ to the Abelian group $U(1)$, the ordinary Clebsch parameters are retrieved. 
We propose two methods to achieve this goal:  

(1) We make use to group elements 
\be
a_\mu =\pa_\mu \theta +\a\pa_\mu \b\rightarrow -i( g^{-1}\pa_\mu g + \a h^{-1}\pa_\mu h)\;,
\ee
where $g,h\in {\cal G}$. It is obvious, as expected, that for a 
$G=U(1)$ group the parameters are  the Clebsch ones.

(2) We use auxiliary fields that are in the algebra of the non-Abelian group
\be
a_\mu =\pa_\mu \theta +\a\pa_\mu \b\rightarrow a^a_\mu =\pa_\mu \theta^a + d^a_{bc}\a^b\pa_\mu \b^c  \;,
\ee
\end{itemize}

In the following subsection we adopt these uplifts of the Clebsch coefficients and develop the corresponding  
Lagrangian formulations, equations of motion, Hamiltonian formulations, symmetries and the coupling to 
non-Abelian gauge fields.

Then, in the subsection after that, 
we summarize the approach of \cite{Bistrovic:2002jx} including the generalization to 
flavored fluids. 

After that,  we consider another formulation for a gluonic fluid coupled to Yang-Mills fields, 
that conserves our defined non-Abelian gluon helicities,
with a Lagrangian formulation and the conservation of the energy-momentum tensor, related to the 
Euler equation. Some classical solutions in this set-up are then considered. Finally, we mention an 
attempt at an Euler equation for a "quark" (non-Abelian in the fundamental) fluid, but we find that it is 
somewhat trivial.

\subsection{The Lagrangian formulation}

Using the uplifts discussed above, the Lagrangian densities for the models of non-Abelian relativistic fluid read
\be\label{Lagna_1}
{\cal L}_1=-\Tr[ j^\mu (g^{-1}\pa_\mu g + \a h^{-1}\pa_\mu h)]   -f(j)\;.
\ee
\be\label{Lagna_2}
{\cal L}_2=- j^\mu_a (\pa_\mu \theta^a + {d^a}_{bc}\a^b\pa_\mu \b^c)   -f(j)\;.
\ee

Notice that these Lagrangian densities  describe a non-Abelian fluid even in the absence of a coupling 
to non-Abelian gauge fields, in a similar way 
to what we have done above in the Abelian case. 
This is different than the approach of \cite{Bistrovic:2002jx}, where 
the non-Abelian fluid was defined only when the fluid couples to the non-Abelian gauge fields.

\subsubsection { The dimensions of the fields}

The passage to non-abelian fluid does not change the discussion of the dimensions of the fields 
$a_\mu,j_\mu, \rho$ and $ \vec v$. The only new ingredients  in (1) are the group elements $g$ and $h$,
which obviously have zero dimensions, $[g]=[h]=0$. 
In (2) $\theta^a, \a^a$ and $\b^a$ also have zero dimensions.

\subsubsection{The equations of motion for the formulation (1)}

\begin{itemize}
\item
The variation with respect to $g\in {\cal G}$ reads
\bea
&&0=\Tr[(-g^{-1}\delta g g^{-1} \pa_\mu g + g^{-1} \pa_\mu \delta g )j^\mu]\rightarrow\cr
&&\Tr\left[-\left\{g^{-1}\delta g \left(\pa_\mu j^\mu + 
[ g^{-1} \pa_\mu g, j^\mu]\right)\right\}\right]=0 \;,
\eea
which implies that 
\be
{D_g}_\mu j^\mu \equiv \pa_\mu j^\mu + [ A^g_\mu, j^\mu]\equiv\pa_\mu j^\mu + 
[ g^{-1} \pa_\mu g, j^\mu]=0.\label{nonabcont}
\ee

This is a non-Abelian generalization of the continuity equation. As it stands it depends on the auxiliary field 
$A^g_\mu$.
\item
The equation of motion associated with the variation of $\a$ is
\be
\Tr[h^{-1}\pa_\mu h  j^\mu]= 0 \;,\label{aNA}
\ee
which is a generalization of (\ref{varab}) of the form
\be
\Tr[\rho^a T^a (h^{-1}\pa_t h+  \vec v\cdot h^{-1}\vec{\nabla} h)]=0.
\ee

\item
In a similar manner to the variation of $g$,  the variation of $h$ yields
\be
\a{D_h}_\mu j^\mu + j^\mu \pa_\mu \a \equiv \a\left(\pa_\mu j^\mu + [A^h_\mu ,j^\mu] \right)
+ j^\mu \pa_\mu \a= 0.\label{AhNA}
\ee
\item
Variation with respect to $j^\mu$ yields
\be
a_\mu = g^{-1}\pa_\mu g + \a h^{-1}\pa_\mu h  =-\frac{u_\mu}{c^2}f'(j)=-\frac{j_\mu}{ j c^2}f'(j).\label{eqnjrNA}
\ee
\item
To get the ``non-Abelian Euler equation'' we follow similar steps as those used to derive (\ref{eqnjr}). 
However, now we want this equation to be covariant, or at least invariant, under ${\cal G}$. Indeed, we 
would like the equation to describe, ideally, $N^2$ degrees of freedom, coupled to Yang-Mills fields. 
And we know that the coupling to Yang-Mills will be covariant, or at least invariant: it should be 
the "Lorentz" force $j^\mu F_{\mu\nu}$ (or, at least, its trace).
We thus again take the curl of the last equation, but now we also add the commutator, to get the 
field strength $F_{(a)\mu\nu}$.
From the left-hand side of the equation (\ref{eqnjrNA}), we get
\bea
\pa_\mu a_\nu-\pa_\nu a_\mu &=& [A^g_\nu,A^g_\mu] + \a [A^h_\nu,A^h_\mu] -(\pa_\nu\a) A^h_\mu 
+ (\pa_\mu\a) A^h_\nu\cr
[a_\mu,a_\nu]&=&[A^g_\mu,A^g_\nu]+\a^2[A^h_\mu,A^h_\nu]+ \a\left([A_\mu^g,A_\nu^h]
-[A_\nu^g,A_\mu^h]\right)
\Rightarrow\cr
F_{(a)\mu\nu}&=&\a(\a-1)[A_\mu^h,A_\nu^h]+\a\left([A_\mu^g,A_\nu^h]-[A_\nu^g,A_\mu^h]\right)
+(\d_\mu\a)A_\nu^h-(\d_\nu\a)A_\mu^h.\cr
&&
\eea
We now multiply this expression with $j^\mu$ and use the relations (\ref{nonabcont}), 
(\ref{aNA}) and (\ref{AhNA}), 
to eliminate $(\d_\mu \a)$, obtaining
\bea
j^\mu F_{(a)\mu\nu}&=&\a(\a-1)j^\mu[A_\mu^h,A_\nu^h]+\a j^\mu\left([A_\mu^g,A_\nu^h]
-[A_\nu^g,A_\mu^h]\right)
\cr
&&+\a[A_\mu^g-A_\mu^h,j^\mu]A_\nu^h-(\d_\nu \a)j^\mu A_\mu^h.
\eea

Unfortunately, there is no way to make this vanish, as it did in the Abelian case (we can, in fact, check that 
the above vanishes in the Abelian case, since commutators vanish and $j^\mu A_\mu^h\rightarrow 
j^\mu \d_\mu \b=0$ by the $\a$ equation of motion). That wouldn't be a problem, a priori, except for the fact 
that we cannot rewrite this in terms of the 4-velocity $u^\mu$ and $j$, it is only a function of the "Clebsch
parametrization" $A_\mu^g, A_\mu^h, \a$. So that does not look like extra terms in a non-Abelian Euler 
equation. Moreover, since it is written only in terms of the Clebsch parameters, but not in terms of the 
velocity and density, 
it is clear that the equation will not preserve the helicities we introduced (which {\em are} written 
only in terms of the velocity and density). 

We might think that taking the trace of the above could help (then we would, at least, get one equation 
right), but it does not. We can now eliminate the term with $\d_\nu\a$, since $\Tr[j^\mu A_\mu^h]=0$ 
by the $\a$ equation of motion, and after some manipulations inside the trace, the rest of the terms 
become
\be
\Tr[j^\mu F_{(a)\mu\nu}]=-\a\Tr\left([A_\nu^h,j^\mu](A_\mu^g+A_\mu^h)\right)+\a(\a-1)\Tr\left(j^\mu[A_\mu^h,
A_\nu^h]\right)\;,
\ee
which is still nonzero, and still not expressible in terms of $u^\mu$ and $j^\mu$ only. Only the 
term $A_\mu^g+A_\mu^h=a_\mu=-u^\mu f'(j)/c^2$ is, the rest are not. 

Finally, the curl plus commutator of the right-hand side of the equation (\ref{eqnjrNA}), multiplied by $j^\mu$, 
which is what we hoped would be the true non-Abelian Euler equation (without extra terms), is again
\be
\frac{j}{c^2}\left\{ u^\mu\d_\mu[u_\nu f'(j)]-\d_\nu f' (j)\right\}+\frac{j}{c^4}[f'(j)]^2u^\mu[u_\mu,u_\nu]\;,
\ee
except, of course, this is now a non-Abelian formula.

\end{itemize}

\subsubsection{The equations of motion for the formulation (2)}

\begin{itemize}
\item
The variation with respect to $\theta^a$ yields, like in the abelian case, a conservation of the 
current with ordinary, not a covariant, derivative,
\be
\pa^\mu j^a_\mu=0.
\ee
\item
From the variation with respect to $\a^a$ we get 
\be
{d^a}_{bc}(\pa^\mu \b^b) j_\mu^c=0.
\ee
\item

In a similar manner, the variation with respect to $\b^a$ yields
\be
{d^a}_{bc}\pa^\mu (  \a^b j_\mu^c)=0.
\ee
\item

From the variation with respect to the current $j^\mu_a$, one finds
\be
a_\mu^a = \pa_\mu \theta^a + {d^a}_{bc}\a^b\pa_\mu \b^c =- \frac{j^a_\mu}{ j c^2}f'(j).\label{eqnjrNA2}
\ee
\item

As before we take now the curl of the last equation,
\be
\pa_\mu a^a_\nu - \pa_\nu a^a_\mu= {d^a}_{bc}(\pa_\mu \a^b\pa_\nu \b^c-\pa_\nu \a^b\pa_\mu \b^c)=
\frac{1}{c^2}(\pa_\mu(u^a_\nu f'(j))-\pa_\nu(u_\mu^a f'(j))) .\label{nonabcurl}
\ee

Upon using the equations of motion  derived above, associated with the variations of 
$\theta^a,\a^a$ and $\b^a$, we have that 
\be
{d^a}_{bc}(\pa_\mu \a^b\pa_\nu \b^c-\pa_\nu \a^b\pa_\mu \b^c) j^\mu_a=0\;,
\ee 
where we see that the index $a$ is summed over (as is, of course, the Lorentz index $\mu$), which means 
that this is a single equation (not a matrix equation, i.e., $N^2$ equations). Thus, in order to be able to 
use it, we must also multiply the equation (\ref{nonabcurl}) with $j_a^\mu$ and sum over $a$. 

Moreover, as in the case of the formulation $(1)$, we must consider a covariant equation, also 
because later we will want to couple to Yang-Mills fields, and the coupling in the Euler equation 
must be in terms of the covariant $F_{\mu\nu}^a$. Therefore, we must again add the commutator 
$[a_\mu, a_\nu]$, also multiplied with $j^\mu_a$. But in the commutator, we can replace the same equation
(\ref{eqnjrNA2}), and so ignore it.

Thus, we get the non-Abelian Euler equation that reads
\be
j^\mu_a\pa_\mu\left[j^a_\nu \frac{f'(j)}{j}\right] - \frac12 \pa_\nu(j^2)\frac{f'(j)}{j}
- j^2\pa_\nu\left[ \frac{f'(j)}{j}\right]=0.\label{relEulerna}
\ee
In this case, we did obtain a non-Abelian Euler equation, but it is not a matrix equation (it is a single equation, 
not $N^2$ equations, like we want).

\end{itemize} 

\subsubsection{Hamiltonian formulations}

In analogy to the  Abelian case, for the case  (1),  
only $g$ and $h$ (but not $\a$ or $j_\mu^a$) have nontrivial conjugate momenta, 
$\pi_g^a=-(j^0 g^{-1})^a$ and $\pi_h^a=-\a(j^0h^{-1})^a$, so the Hamiltonian density is 
\be
{\cal H}=-\Tr[j^0(g^{-1}\d_0 g+\a h^{-1}\d_0 h)]-{\cal L}
=\Tr\left[\vec{j}\cdot\left(g^{-1}\vec{\nabla}g+\a h^{-1}\vec{\nabla}h\right)\right]+f(j).
\ee

For the Lagrangian density (2), $\theta^a$ and $\b^a$ have conjugate momenta
\be
\pi_{\theta^a} = -j_0^a=-\rho^a\qquad \pi_{\b^a}=- {d^a}_{bc}\a^bj_0^c=- {d^a}_{bc}\a^b\rho^c.
\ee

The corresponding Hamiltonian density is
\be
{\cal H}=-\rho^a\dot\theta_a- {d^a}_{bc} \a^b\rho^c\dot \b_a-{\cal L}
= \vec j^a( \vec\nabla \theta_a+ d_{abc} \a^b\vec\nabla \b^c) + f(j).
\ee

\subsubsection{The symmetries}

\begin{itemize}

\item As in the Abelian case, we have Lorentz symmetry, parity and time-reversal symmetry, where 
in the case (1) we have
$g(-t)=-g(t)$, $h(-t)=-h(t)$ and $g(-\vec{x})=-g(\vec{x})$, $h(-\vec{x})=-h(\vec{x})$, and in the case (2)
we have the obvious non-Abelian generalization of the Abelian case.
\item In the case (1), we also 
have ${\cal G}$ symmetry invariance, separately for $g$ and for $h$, so we have really 
${\cal G}\times {\cal G}$ invariance. 

\item Translational invariance, giving the Noether current = energy-momentum tensor. In the case (1), 
\bea
T_{\mu\nu}&=&-\sum_{\phi_q=g,h} \frac{\d {\cal L}}{\d \d^\mu \phi_q}\d_\nu \phi_q +\eta_{\mu\nu}{\cal L}\cr
&=& \Tr\left[j_\mu \left(g^{-1}\d_\nu g+\a h^{-1}\d_\nu h\right)\right]
-\eta_{\mu\nu}\left\{\Tr\left[j^\rho\left(g^{-1}\d_\rho g+\a h^{-1}\d_\rho  h\right)\right]+f(j)\right\}\;,\cr
&&
\eea
whereas in the case (2), we have 
\bea
T_{\mu\nu}&=&-\sum_{\phi_q=\theta,\b} \frac{\d {\cal L}}{\d \d^\mu \phi_q}\d_\nu \phi_q +\eta_{\mu\nu}{\cal L}\cr
&=& \left[j_\mu^a \left(\d_\nu \theta^a+d_{abc} \a^b\d_\nu \b^c\right)\right]
-\eta_{\mu\nu}\left\{j^\rho_a\left(\d_\rho \theta^a+d_{abc}\a^b \d_\rho  \b^c\right)+f(j)\right\}\;,\cr
&&
\eea

\end{itemize}

\subsubsection{The fluid coupled to non-Abelian gauge fields}

So far we have discussed a non-Abelian fluid, but without coupling it to  non-Abelian gauge fields. 
This was the analog of ordinary Abelian fluid not coupled to external electromagnetic 
fields.  Turning on a coupling to 
a background of non-Abelian gauge fields, one has to add to the auxiliary  non-Abelian gauge field $a_\mu$ 
a genuine gauge field $A_\mu$. 
It is important to emphasize that similarly to the coupling to the electromagnetic fields, here also we 
do not include  the backreaction of the fluid on the non-Abelian gauge fields. The incorporation of the interaction 
is thus done by the replacement $a_\mu\rightarrow a_\mu+A_\mu$, which in the case (1) becomes
\be
a_\mu \rightarrow a_\mu + A_\mu= -i(g^{-1}\pa_\mu g + \a h^{-1}\pa_\mu h)   + A_\mu^a T^a\;,
\ee
so that the Lagrangian density reads
\be\label{LagnaA}
{\cal L}=-\Tr\left[ j^\mu (g^{-1}\pa_\mu g + \a h^{-1}\pa_\mu h +A_\mu)\right]   -f(j)\;,
\ee
and in the case (2) becomes
\be
a_\mu^a\rightarrow a_\mu^a+A_\mu^a=\d_\mu \theta^a +{d^a}_{bc}\a^b \d_\mu \b^c +A_\mu^a\;,
\ee
and the Lagrangian becomes
\be
{\cal L}=-j^\mu_a (\d_\mu \theta^a+{d^a}_{bc}\a^b \d_\mu \b^c+A_\mu^a)-f(j).
\ee

The "Euler" equation will now have the source
\be
j^\mu F_{\mu\nu}\;,
\ee
as a matrix source in the case (1), 
which is the correct "Lorentz" force coupling to Yang-Mills fields. 
Except, of course, we saw that the resulting Euler equation is not very useful. And the source 
$j^\mu_a F^a_{\mu\nu}$ in the case (2), which is good, but it gives only one equation, as we said.


\subsection{The approach of \cite{Bistrovic:2002jx}  for a fluid coupled to non-Abelian gauge fields }

A different uplift of the Clebsh coefficients  that leads to a different Lagrangian formulation was proposed in 
\cite{Bistrovic:2002jx}.   This paper considers a non-Abelian current $J^\mu_a(t,\vec{r})$ that, however,
splits into a standard Abelian current $j^\mu(t,\vec{r})=\rho(t,\vec{r}) u^\mu(t,\vec{r})$, with an Abelian 
4-velocity $u^\mu$, 
together with a non-Abelian charge $Q_a(t,\vec{r})$, obeying a Wong equation along the fluid particle worldlines,  
\be
\frac{d{\cal Q}_a(\tau)}{d\tau}+f_{abc}\frac{dX^\mu(\tau)}{d\tau}A_\mu^b(X^\nu(\tau)){\cal Q}_c(\tau)=0\;,
\ee
which can be expressed as the covariant conservation 
\be
j^\mu (D_\mu Q)_a=0\;,
\ee
and is the equivalent of the $\a_a,\b^a$ equations of motion in the Abelian case.
If it is satisfied, this, together with the conservation $Q\d^\mu j_\mu=0$, gives
\be
D_\mu(j^\mu Q)=D_\mu J^\mu=0\;.\label{nonabcurcons}
\ee

The nonrelativistic non-Abelian Lagrangian is written in terms of a an arbitrary group element $g\in {\cal G}$,
taking the role of the parameters of the Clebsch parametrization ($\theta,\a,\b$), 
and a fixed Lie algebra element $T_0$ (so $g=\exp [i\a^a T_a]$), as\footnote{Note that  in \cite{Jackiw:2000cd}
a related version of a non-Abelian Clebsch parametrization was considered, and Chern-Simons
terms were also analyzed, but in this case still we had a fix Lie algebra element projection, and the 
velocity $u^\mu$ and current $j^\mu$ were still Abelian. Further, in \cite{Nair:2020kjg}, various such 
Chern-Simons terms were added to the action as topological terms, 
but were not considered as non-Abelian helicities, as we do.}
\be
{\cal L}=j^\mu 2\Tr[T_0 g^{-1}D_\mu g]-f(j)+{\cal L}(A_\mu)\;,\;\; D_\mu=\d_\mu +A_\mu.
\ee

Note the normalization is $\Tr[T^aT_b]=-(1/2)\delta^a_b$. This is invariant under the gauge group 
$g\in {\cal G}$ as follows:
left action on $g$, $g\rightarrow U \cdot g$, and adjoint action on $A_\mu$, $A_\mu\rightarrow U (A_\mu
+\d_\mu)U^{-1}$. 

If we ignore the $A_\mu$ Lagrangian ${\cal L}(A_\mu)$, we get coupling with {\em external} 
(non-dynamical) electromagnetism, 
just like in the cases studied in \cite{Abanov:2021hio,Nastase:2022aps}.

Then:
\begin{itemize}

\item The equation of motion of $g$ gives the current conservation, (\ref{nonabcurcons}).

\item Varying with respect to $j^\mu$ gives a sort of Bernoulli equation, and then taking $\d_\nu$ on it 
and antisymmetrizing 
in $(\mu,\nu)$, manipulating the result, one obtains the non-Abelian Euler-type equation
\be
\frac{j u^\mu}{c^2}\d_\mu(u_\nu f'(j))-j\d_\nu f'(j)=2\Tr[J^\mu F_{\mu\nu}]=-j^\mu Q_a F^a_{\mu\nu}\;,
\ee
with the right-hand side describing the non-Abelian Lorentz force term, the generalization of the 
Abelian term $j^\mu F_{\mu\nu}$, which for $\nu=i$ gives
\be
j^\mu F_{\mu i}=\rho\left(\vec{E}+\frac{\vec{v}}{c}\times\vec{B}\right)_i\;,
\ee
and the left-hand side the Euler equation multiplied by  $\rho$.

Indeed, the non-relativistic limit is now
\be
\d_t\vec{v}+\vec{v}\cdot\vec{\nabla}\vec{v}+\frac{\vec{\nabla} P}{\rho}=Q^a\left[\vec{E}_a
+\frac{\vec{v}}{c}\times \vec{B}_a\right]\;,
\ee
and $E^i_a\equiv cF_{0i}^a$, $B^i_a\equiv-\frac{c}{2}\epsilon^{ijk}F_{jk}^a$.

\end{itemize}

\subsection{Flavored colored fluid}

One can consider the generalization corresponding to several types of fluid velocity, which 
is by choosing several directions in the Lie algebra, $T_f$, such that the $T_f$'s 
commute among each other, i.e., they belong to the Cartan subalgebra, 
and equal in number the rank of the group
(so $N$ for $U(N)$). This is then somewhat similar to the case of the quark fluid that we considered in 
the previous section.
Each component $f$ has an associated velocity, density and current, so 
$u^\mu_f, j_f, j^\mu_f$, and one has 
\bea
J^\mu&=&\sum_{f=1}^N Q_{f} j^\mu _{f}=\sum_{f=1}^N\sum_{a=1}^{N^2} Q_{af }T^aj^\mu_{f}\cr
Q_{f}& =& Q_{af}T^a=gT_{f} g^{-1}.
\eea

The Lagrangian is a sum over $f$, specifically
\be
{\cal L}=\sum_{f=1}^Nj^\mu_{f}2\Tr[T_{f}g^{-1}D_\mu g]-f(\{j_{f}\})+{\cal L}(A_\mu ^a)\;,
\ee
with the Wong equation  
\be
\sum_{f}j^\mu_{f}D_\mu Q_{f}=0.
\ee

One can write would-be Euler equations for each $f$ channel, so 
\be
\frac{j_{f}u^\mu_{f}}{c^2}\d_\mu\left(u^\nu _{f}\frac{\d f}{\d j_{f}}\right)-j_{f}\d^\nu \frac{\d f}{\d
j_{f}}=2\Tr\left[j^\mu_{f}(D_\mu Q_{f})(D^\nu g)g^{-1}\right]+2\Tr[j_{\mu f}Q_{f}F^{\mu\nu}].
\ee

However, as we can easily see, {\em only summing over $f$ do we get 
something like the Euler equation coupling
on the right-hand side}, obtaining
\be
\sum_{f}\left\{\frac{j_{ f}u^\mu_{f}}{c^2}\d_\mu\left(u^\nu _{f}\frac{\d f}{\d j_{f}}\right)
-j_{f}\d^\nu \frac{\d f}{\d j_{f}}\right\}=-\sum_fj_{\mu f}Q_{af}F^{a\mu\nu}\;,
\ee
and the right-hand side becomes, for $\nu=i$, the Lorentz force
\be
\sum_{f}\rho_{f} Q_{af}\left(\vec{E}_a+\frac{\vec{v}_{f}}{c}\times \vec{B}_a\right)_i.
\ee

So, this summed over Euler equation in the nonrelativistic limit becomes (with $f=\sum_{f}j_{f}c^2-V$)
\be
\sum_{f}\rho_{f}\left[\d_t\vec{v}_{f}+\vec{v}_{f}\cdot\vec{\nabla}\vec{v}_{f}\right]+\vec{\nabla}P
=\sum_f\rho_{f} Q_{af}\left(\vec{E}_a+\frac{\vec{v}_{f}}{c}\times \vec{B}_a\right)\;,\label{nAbEu2}
\ee
but {\bf the non-summed Euler equation doesn't have a simple answer 
on the right-hand side, so cannot be understood as a 
true Euler equation}. 

In conclusion, this case only gives {\em one} sensible Euler equation, and not one for each fluid component
(and in any case the fluid components were quark-like, not gluon-like).

\subsection{Nonabelian gluonic fluid  coupled to Yang-Mills fields}

Different non-Abelian fluid Euler equations were considered, for instance, in \cite{Bistrovic:2002jx} and 
\cite{Cheung:2020djz}. However, as we said, our goal is to write non-Abelian generalizations of the 
Euler equation that can conserve the helicities we 
defined, and perhaps even be obtained from a Lagrangian. 
Also, as opposed to the case in  \cite{Bistrovic:2002jx}, we will obtain 
Euler equations for each adjoint index $a$, and not 
only a single one.

For the (effective) gluonic fluid, interacting with Yang-Mills fields, we then propose the generalized Euler equation 
\be
\left[\d_t\vec{v}_{a}+\vec{v}_{a}\cdot\vec{\nabla}\vec{v}_{a}\right]+\vec{\nabla}\mu_a
= Q_{a}\left(\vec{E}_a+\frac{\vec{v}_{a}}{c}\times \vec{B}_a\right)\;,\label{gluonEuler}
\ee
{\em with no sum over $a$}, and with (again, without sum over $a$, unless specifically written)
\be
\Pi_\mu=\sum_a \Pi^a_\mu T_a\;,\;\;
\vec{\Pi}_a=m_a\vec{v}_a+Q_a\vec{A}_a\;,\;\;
\Pi_{0a}=-\mu_a-m_a\frac{\vec{v}^2_a}{2}+A_{0a}.
\ee

Using the 4-dimensional notation (even though we have a nonrelativistic theory), as done in \cite{Abanov:2021hio} in the 
Abelian case, for $\Pi_\mu$ and the 4-current
\be
j_\mu^ a=\rho^ a u_\mu^ a\;,
\ee
the gluonic Euler equation (\ref{gluonEuler}) implies (with no sum over $a$)
\be
j^{a\mu}\Omega_{\mu\nu}^ a=0\;.
\ee

Note that $\Omega_{0i}^ a=E_i^ a, \Omega_{jk}^ a=\epsilon_{ijk}B_i^ a$.

This 4-dimensional form of the non-Abelian Euler equation can be used, 
like in the Abelian case in \cite{Abanov:2021hio}, 
to prove the conservation of the helicities we have defined. 

Indeed, defining
\be
\Pi=\sum_a\Pi^ a T_a\;,\;\; \Omega=\sum_a\Omega^ aT_a\;,\;\; \Omega=d\Pi+\Pi \wedge \Pi\;, 
\ee
the helicity densities
\be
h^a=\Tr\left[T^ a\left(\Pi \wedge d\Pi +\frac{2}{3}\Pi\wedge \Pi\wedge \Pi\right)\right]\;,
\ee
obey
\be
dh^ a=\Tr\left[T^ a\left(\Omega\wedge \Omega\right)\right].
\ee

Then, by a generalization of the argument in the Abelian case, we have $\Omega^ a\wedge \Omega^ b=0$, since
$\Omega^a \wedge \Omega^ b$ is a 4-form, thus proportional to the volume form. 
But, if we contract with $j^{a \mu}$, 
so (with no sum over $a$) $j^ {a\mu}\Omega^ a_{[\mu\nu}\Omega^ b_{\rho\sigma]}=0$, 
we get zero because of the Euler equation above. 
But that is only possible if the original proportionality constant was zero, 
so $\Omega^ a\wedge \Omega^ b$ was zero to begin with.
Explicitly, with indices, we have the identity
\be
\epsilon^ {\a\nu \lambda \rho}4\Omega_{[\mu\nu}^ a\Omega^ b_{\lambda\rho]}
=\delta_\mu^\a \epsilon^ {\tau\nu \lambda\rho}
\Omega^ a_{\tau\nu}\Omega^ b_{\lambda\rho}\;,
\ee
which is correct, since both $\a $ and $\mu$ must be different than $\nu\lambda\rho$, 
hence must be equal, and the rest follows.

But then, multiplying with $j^{a\mu}$ gives zero because of the gluonic Euler equation, 
while on the right-hand side 
we get $j^ {a\a}(\Omega^ a\wedge \Omega^ b)$, so $\Omega^ a\wedge \Omega^ b=0$. But then
\be
dh^ a=\sum_{b,c}\Tr[T^ a T^ b T^ c]\Omega^ b \wedge \Omega^c=0\;,
\ee
so all the $N^ 2$ ${{\cal H}^{NA}}_{mm}^a$ (as well as the singlet one ${\cal H}^{NA}_{\rm total}$, 
by the same argument) helicities are conserved!

\subsection{Lagrangian formulation  and dynamical Yang-Mills fields}

Now we want to obtain the equations (\ref{gluonEuler}), together with the non-Abelian 
current conservation equation, 
from a Lagrangian. 

Unlike the case in \cite{Bistrovic:2002jx}, we consider $j^\mu=j^{\mu a}T_a$ a fully nonabelian current, 
without a split into $j^\mu$ and $Q^a$, so that we have a chance to obtain the Euler equation with an index $a$.

The action we try is then (note that there is no $\a h^{-1}\d_\mu h$ term, and $g^{-1}\d_\mu g +A_\mu$ 
is replaced by $g^{-1}D_\mu g$, with respect to the previous attempt)
\be
S=\int d^4x \left\{\Tr[j^\mu g^{-1}D_\mu g]-\sum_a\sqrt{-j^{\mu a}j_\mu ^a}
-V\left(\sum_a\sqrt{-j^{\mu a}j_\mu ^a}\right)
+{\cal L}(A_\mu^a)\right\}\;,
\ee
where $V$ is a potential, depending on the nonrelativistic non-Abelian 
generalization of the $V(j)$ in the Abelian case, 
\bea
D_\mu g&=& \d_\mu g+A_\mu g\cr
j^{\mu a}&=& \rho_0^a u^{\mu a}=\rho^a\sqrt{1-\vec{v}_a^2}\frac{dx^{\mu a}}{d\tau}\;,
\eea
and $u^{\mu a}$ is the 4-velocity of the fluid particle of $a$ type.

Note that 
\be
-j^{a\mu} j^a_\mu =\rho_0^a\Rightarrow \frac{d}{d j^{\mu a}}\sqrt{-j^{\mu a}j^a_\mu}=u^a_\mu\;.
\ee

Then we obtain:

-the $j_\mu^a$ equation of motion as a simpler non-Abelian generalization of the Clebsch parametrization of the 
velocity (\ref{vClebsch}),
\be
u_\mu^a(1+V')=(g^{-1}D_\mu g)^a.\label{uaeq}
\ee

-the continuity equation is obtained from the variation of the action with respect to $g$, {\em after which 
we fix the gauge $g=1$,} which gives 
\be
(D_\mu j^\mu)^a=0.\label{contYM}
\ee

-the $A_\mu^a$ equation of motion, if we consider the gauge fields as dynamical, is 
\be
(D_\nu F^{\nu\mu})^a=j^{\mu a}.
\ee
If we ignore ${\cal L}(A_\mu^a)$, the gauge fields are treated as external. 

Taking $\d_\mu$ (\ref{uaeq})$_\nu$ $-(\mu\leftrightarrow \nu)$ gives 
\be
\d_\mu (g^{-1}D_\nu g)^a-\d_\nu(g^{-1}D_\mu g)^a
=(\d_\mu u_\nu^a-\d_\nu u_\mu^a)(1+V')+u_\mu^a \d_\nu V'-u_\nu^a\d_\mu V'.
\label{gua}
\ee

Multiplying by $\rho_0^a u^{\mu a}$, ignoring the $V'$ in $(1+V')$ and $u^{\mu a}\d_\mu V'$ as being subleading
in the nonrelativistic limit (otherwise we keep it there), 
and since $u^{\mu a}\d_\nu u_\mu^a=\d_\nu(u^{\mu a}u^a_\mu)=\d_\nu(-1)=0$, the 
right-hand side gives approximately
\be
\rho_0^a u^{\mu a}\d_\mu u_\nu^a-\rho_0^a\d_\nu V'\;,
\ee
which for $\nu=i$ and in the nonrelativistic limit gives the left-hand side of the Euler equation (with an index $a$). 

The left-hand side of (\ref{gua}), contracted with $T_a$ and taking the trace, gives 
\bea
\d_\mu\Tr[g^{-1}D_\nu g]-\d_\nu \Tr[g^{-1}D_\mu g]&=&\Tr[D_\mu(g^{-1}D_\nu g)-D_\nu(g^{-1}D_\mu g)]\cr
&=&\Tr\left\{g^{-1}[D_\mu,D_\nu ]g-[g^{-1}D_\mu g,g^{-1}D_\nu g]\right\}\cr
&=& \Tr\left\{g^{-1}F_{\mu\nu}g-[g^{-1}D_\mu g,g^{-1}D_\nu g]\right\}.
\eea

But now we peel off the $T_a$ with the trace (so the $(T_a)_{ji}$), 
by multiplication with $(T^a)^{kl}$ and completeness of the 
$T_a$ matrices, we remember that we have (\ref{uaeq}), so we get 
\be
(g^{-1}F_{\mu\nu}g)^a-\sum_{b,c}{f^a}_{bc}u^b_\mu u^c_\nu\;,
\ee
and this was contracted with $\rho_0^a u^{\mu a}$, in order to (hope to) 
give  the right-hand side of the Euler equation. 

But we saw that the gauge $g=1$ is needed (it was needed to obtain 
the correct continuity equation (\ref{contYM})), 
and then, putting both sides together and $g=1$, the Euler equation {\em in the nonrelativistic limit} is now 
\be
\rho_0^a u^{\mu a}\d_\mu u_\nu ^a-\rho_0^a \d_\nu V'
=\rho_0^aF_{\mu\nu}^a u^{\mu a}-\rho_0^a \sum_{b,c}f_{abc}u^{\mu a}u_\mu^b u_\nu^c
=\rho_0^a F_{\mu\nu}^a u^{\mu a}\;,
\ee
since the last term vanishes ($u^{\mu a}u_\mu^b$ is symmetric in $(ab)$, but $f_{abc}$ is antisymmetric). 

This is indeed just the Euler equation we wanted.

\subsection{Belinfante energy-momentum tensor and the Euler equation}

The standard Euler equation is understood as the nonrelativistic limit of the conservation equation for the 
energy-momentum tensor of a fluid, $\d^\nu T_{\mu\nu}=0$, where 
\be
T^ {\mu\nu}=\rho u^\mu u^\nu+P(\eta^{\mu\nu}+u^\mu u^\nu).\label{tmunufl}
\ee
So it is natural to ask: can the same be true in the non-Abelian 
case presented here?

But first, we want to understand how this works in the Abelian case, not from the $T_{\mu\nu}$ of a fluid, 
as we did in the previous fluid cases, but from 
the {\em Belinfante} energy-momentum tensor (that should therefore 
give the same result as the Noether current calculation for the Abelian cases
in section 4) associated with the nonrelativistic Abelian Lagrangian for the fluid, 
\be
{\cal L}=-j^\mu a_\mu -\sqrt{-j^\mu j_\mu}-V\;.\label{relagrav}
\ee

Coupling it to gravity, we get 
\be
\frac{{\cal L}}{\sqrt{-g}}=-j^\mu a_\mu -\sqrt{-j^\mu j^\nu g_{\mu\nu}}-V\left(\sqrt{-j^\mu j^\nu g_{\mu\nu}}\right)\;,
\ee
since $A_\mu$ and $j^\mu$ are the fundamental fields (note that for a 
particle $j^\mu =q\frac{dx^\mu}{d\tau }\delta^4(x-x(\tau))$),
and where 
\bea
j^\mu &=&\rho \sqrt{1-v^2}u^\mu=\rho_0\frac{dx^\mu}{d\tau}\cr
u^\mu&=&\frac{dx^\mu}{d\tau}=\frac{1}{\sqrt{1-v^2}}(1,\vec{v})=\frac{\tilde u^\mu}{\sqrt{1-v^2}}\;,\;\;
\tilde u^\mu=\frac{dx^\mu}{dt}.
\eea

Then the Belinfante energy-momentum tensor is obtained as 
\be
T^{\mu\nu}=-\frac{j^\mu j^\nu}{\sqrt{-j^\rho j_\rho}}(1+V')+\eta^{\mu\nu}{\cal L}=\rho \tilde u^\mu \tilde u^\nu(1+V')
+\eta^{\mu\nu}{\cal L}\;,
\ee
and the on-shell Lagrangian becomes (since $a_\mu=u_\mu (1+V')$) 
${\cal L}\simeq \rho V' -V\simeq \rho V'$ (assuming
we can neglect $V$), meaning we get 
\be
T^{\mu\nu}=\rho\tilde u^\mu \tilde u^\nu +\left(\eta^{\mu\nu}+\tilde u^\mu \tilde u^\nu\right)(\rho V')\;,
\ee
which exactly matches the energy-momentum tensor of the fluid, (\ref{tmunufl}), 
which means that it will lead to the 
correct Euler equation
($\d^\mu T_{\mu 0}=0$ 
gives the continuity equation, and $\d^\mu T_{\mu i}=0$ gives a combination of the 
continuity equation and the Euler equation).

In the non-Abelian case then, we again couple the action (for simplicity, 
without the kinetic term for $A_\mu^a$, though that 
can be introduced as well) to gravity, obtaining 
\be
S=\int d^4x  \left\{\Tr[j^\mu g^{-1}D_\mu g]-\sum_a\sqrt{-j^{\mu a}j^{\nu a}
g_{\mu\nu}}-V\left(\sum_a\sqrt{-j^{\mu a}j^{\nu a}g_{\mu\nu}}
\right)\right\}\;,
\ee
leading to the Belinfante energy-momentum tensor
\be
T^{\mu\nu}=-\sum_a\frac{j^{\mu a}j^{\nu a}}{\sqrt{-j^{\rho a} j^a_\rho}}+\eta^{\mu\nu}{\cal L}=\sum_a
\rho^a \tilde u^{a\mu} \tilde u^{a\nu}+\eta^{\mu\nu}{\cal L}\;,
\ee
and on-shell (at $A_\mu=0$, in the $g=1$ gauge) ${\cal L}\simeq \sum_a\rho^a V'$, so we get
\be
T^{\mu\nu}\simeq \rho  \sum_a \tilde u^{a\mu}\tilde u^{a\nu}
+\sum_a\left(\eta^{\mu\nu}+\tilde u^{a\mu}\tilde u^{a\nu}\right)
\rho^a V'\equiv \sum_a T^{\mu\nu a}.
\ee

We see then that we could {\em define} a $T_{\mu\nu}^a$ for each $a$ index, 
and then the conservation of each $T_{\mu\nu}^a$ will give the 
left-hand sides of the Euler equation with index $a$.

We could also keep the Yang-Mills fields $A_\mu^a$, 
both the coupling and the dynamical terms, and the result would generalize
as well.

\subsection{Null fields for classical solutions}

Here we review a formalism that has been used to find classical 
solutions of electromagnetism, e.g. in \cite{Hoyos:2015bxa}, 
fluids \cite{Alves:2017ggb,Alves:2017zjt} 
and electromagnetism coupled to fluids \cite{Nastase:2022aps}.

The electromagnetic fields $\vec E$ and $\vec B$ can be combined into a    complex vector, the so-called
Riemann-Silberstein vector,
\be
\vec{F}=\vec{E}+i\vec{B}\;.
\ee

The Maxwell's equations of motion in vacuum, in terms of $\vec{F}$, are 
\be
\vec{\nabla}\cdot\vec{F}=0\;;\;
\d_t\vec{F}+i\vec{\nabla}\times \vec{F}=0. \label{MxF}
\ee

One can then define null fields by
\be\label{nullEB}
F^2=0 \qquad \leftrightarrow \qquad\vec E\cdot \vec B=0 \qquad E^2-B^2=0.
\ee

A useful parametrization of the null fields  is in terms of  the Bateman ansatz \cite{Bateman:1915},
\be
\vec{F}=\vec{\nabla}\a\times\vec{\nabla}\b\;,\label{bateman}
\ee
in terms of the two complex scalar fields $\a,\b\in\mathbb{C}$. 

In terms of electric and magnetic fields, the ansatz is 
\bea
E^i&=& \epsilon^{ijk} \left(\pa_j\aR\pa_k\bR- \pa_j\aI\pa_k\bI\right)\cr
B^i&=& \epsilon^{ikj} \left(\pa_j\aR\pa_k\bI+ \pa_j\aI\pa_k\bR\right)\;,\label{compEB}
\eea
where the indices $I$ and $R$ refer to the imaginary and real parts, respectively.

The first  Maxwell equation  in (\ref{MxF})  is  trivially satisfied by the Bateman ansatz (\ref{bateman}) 
and the second Maxwell equation takes the form 
\be
i\nabla\times (\pa_t\alpha\nabla\beta-\pa_t\beta\nabla\alpha)=\nabla\times \vec F\;,\label{FMx}
\ee
which is satisfied if one has
\be
i(\pa_t\alpha\nabla\beta-\pa_t\beta\nabla\alpha)=\vec F.
\ee

In fact in a more general solution the right-hand side could be 
$\vec F +\vec G$, where $\nabla\times \vec G=0$, but $\vec G$ can be time dependent.

We note that {\em on-shell}, we can write 
\be
A_\mu= \frac12{\rm Im}[ \alpha\pa_\mu \beta- \beta\pa_\mu \alpha]\;,
\ee
and upon denoting $\alpha=\alpha_R+ i \alpha_I$ and $\beta=\beta_R+ i \beta_I$, the gauge field takes the form

\be\label{AaRIbRI}
A_\mu=\frac12{\rm Im}\left[ \alpha_R\pa_\mu \beta_I + \alpha_I\pa_\mu
 \beta_R-[(\pa_\mu \alpha_R)\beta_I+ \pa_\mu \alpha_I)\beta_R]\right].
\ee

Solutions of (\ref{FMx}) have necessarily a zero norm, since 
\be\label{null}
F^2=(\pa_t\alpha\nabla\beta-\pa_t\beta\nabla\alpha)(\vec{\nabla}\a\times\vec{\nabla}\b)=0 \;.
\ee

The simplest null electromagnetic field is made up of perpendicular constant electric and magnetic fields, namely
\be\label{basicEM}
\vec E = (E,0,0)\;, \qquad \vec B = (0,B,0)\;,\qquad |E|=|B|.
\ee

In terms of $\a$ and $\b$, for $E=-B=-4$, this solution takes the form
\be\label{Bateconstant}
\a= 2i(t+z)-1\;, \qquad \b=2(x-iy).
\ee

Another class of well-known null electromagnetic solutions are the plane waves
\be
\vec F = (\hat x +i \hat y) e^{i(z-t)}\;,
\ee
which in terms of the Bateman variables read
\be
\alpha= e^{i(z-t)}\;,\qquad \beta = x+ i y.
\ee

The basic knotted solution, the Hopfion solution 
is given by
\be
\a=\frac{A-iz}{A+it}\;,\;\;
\b=\frac{x-iy}{A+it}\;,\;\;
A=\frac{1}{2}(x^2+y^2+z^2-t^2+1).\label{absol}
\ee

The Hopfion is 
characterized by non-trivial helicity ${\cal H}_{mm} =\frac14$.

This solution can be obtained by applying the  special conformal transformation (SCT)
\be
x^\mu \rightarrow \frac{x^\mu -b^\mu x_\sigma x^\sigma}{1- 2 b_\sigma x^\sigma 
+ b_\rho b^\rho x_\sigma x^\sigma}
\ee
on (\ref{Bateconstant}), with $b^\mu =(i,0,0,0)$.
   
From the Maxwell electromagnetism, we can map to null fluids, with $\vec{v}^2=1$, so $u^\mu u_\mu=0$, via
the map
\be
\rho\leftrightarrow \frac{1}{2}(\vec{E}^2+\vec{B}^2);\;\;\;\;
v_i\leftrightarrow\frac{[\vec{E}\times \vec{B}]_i}{\frac{1}{2}(\vec{E}^2+\vec{B}^2)}\;,\label{velmap}
\ee
valid for null electromagnetic fields, $\vec{F}^2=0$.

Then one finds the basic fluid Hopfion (knotted) solution of the Euler equation and continuity equations,
\bea
&&v_x=\frac{2(y+x(t-z))}{1+x^2+y^2+(t-z)^2}\;,\;\;
v_y=\frac{-2(x-y(t-z))}{1+x^2+y^2+(t-z)^2}\;,\cr
&&v_z=\pm \sqrt{1-v_x^2-v_y^2}=\pm \frac{1-x^2-y^2+(t-z)^2}{1+x^2+y^2+(t-z)^2}\;,\cr
&&\rho=\frac{16(1+x^2+y^2+(t-z)^2)^2}{\left(t^4-2t^2(x^2+y^2+z^2-1)+(1+x^2+y^2+z^2)^2\right)^3}\;.\;\;\; \label{rhov}
\eea

\subsection{Classical solutions}

In order to find classical solutions of the non-Abelian Euler fluid coupled to Yang-Mills equations, 
we do the same as in the Abelian case, which was considered in \cite{Nastase:2022aps}.

We can then find two types of solutions:

1) We can use the null fluid - electromagnetism map from \cite{Alves:2017ggb,Alves:2017zjt}
to define the null velocity field for each index $a$,
\be
v^i_a=\frac{\epsilon^{ijk}E^j_a B^k_a}{\vec{B}^2_a}\Rightarrow 
-(\vec{v}_a\times \vec{B}_a)_i=-\epsilon_{ijk}v^j_aB_a^k
=-\frac{\epsilon^{ijk}\epsilon^{jlm}E_l^aB_m^aB_k^a}{\vec{B}^2_a}
=E_i^a\;,
\ee
with no sum over $a$. 

That means that the source of the non-Abelian Euler equation (its right-hand side)
vanishes, so if the fluid $v^i_a$ is a solution of our generalized Euler equation (with index $a$), 
derived from the solution of the Yang-Mills equation, then it is a solution of the 
combined system. 

2) We can, alternatively, consider the YM fields as just {\em external} 
fields, which only need to satisfy the Bianchi identities 
$D_{[\mu}F_{\nu\rho]}=0$, saying that $F_{\mu\nu}^a$ is derived from an $A_\mu^a$.

But then, 
\be
v^i_a=\frac{\epsilon^{ijk}E_a^jB_a^k}{\frac{1}{2}(\vec{E}^2_a+\vec{B}^2)}
\ee
satisfies 
\be
u^\mu_aF_{\mu\nu}^a=0: \vec{v}_a\cdot\vec{E}_a=0\;\;\;\&\;\;\; 
\vec{E}_a+\vec{v}_a\times \vec{B}_a=0.
\ee

Then we replace $F_{\mu\nu}^a$ with 
\be
\Omega_{\mu\nu}^a=F_{\mu\nu}^a+{\cal V}_{\mu\nu}^a\;,
\ee
where 
\be
{\cal V}^a_{0i}=m(\d_0 v_i^a-v_j^a\d_iv^j_a)\;,\;\;\;
{\cal V}^a_{ij}=m(\d_i v_j^a-\d_j v_i^a)\;,
\ee
in the YM  solution $F_{\mu\nu}^a=...$, which means we now have, instead,
\be
u^\mu_a \Omega_{\mu\nu}^a=0\;,
\ee
which is our generalized Euler equation (provided $n^a=n$ is the same). 

But by construction $\d_{[\mu}{\cal V}_{\nu\rho]}^a=0$, and we also have (since it satisfies the YM equations) 
$D_{[\mu}\Omega_{\nu\rho]}^a=0$, which means that the Bianchi identities are satisfied,
\be
D_{[\mu}F_{\nu\rho]}^a=D_{[\mu}\Omega_{\nu\rho]}^a-D_{[\mu}{\cal V}^a_{\nu\rho]}=D_{[\mu}\Omega_{\nu\rho]}
-\d_{[\mu}{\cal V}_{\nu\rho]}^a=0\;,
\ee
as well, as we wanted.

Of course, the most general case is when the we have a solution in the presence of dynamical Yang-Mills fields, 
which means we would need to satisfy, besides the non-Abelian Euler 
equation and Bianchi identities, also
the equation of motion 
\be
(D_\mu F^{\mu\nu})^a=j^{\nu a}.
\ee

But that is beyond what we can do at this time.

\subsection{Attempt at an Euler equation for non-Abelian fluid in the fundamental ("quark")}

For completeness, we describe here also an attempt to construct Euler 
equations for the non-Abelian fluid in the fundamental 
representation (for a "quark" fluid), though we will see that while we 
can preserve the helicities ${\cal H}'_f$ defined in 
(\ref{htot}), there is no interaction between the fluid types, so 
it is not clear if these Euler equations are useful.

We could, indeed, {\em define} the generalized non-Abelian Euler 
equation for fluid velocity with an index $f$ in the 
Cartan subalgebra,
\be
\left[\d_t\vec{v}_{f}+\vec{v}_{f}\cdot\vec{\nabla}\vec{v}_{f}\right]+\vec{\nabla}\mu_f
= \sum_a Q_{af}\left(\vec{E}_a+\frac{\vec{v}_{f}}{c}\times \vec{B}_a\right)\;,\label{nAbEu1}
\ee
with no sum over $f$. We could also consider the case that the sum 
over $a$ runs only over an index $\tilde b$ that 
excludes the Cartan subalgebra.

The Euler equation (\ref{nAbEu1}) preserves the helicity ${\cal H}'_{\rm tot}$ 
defined in (\ref{htot}), which is seen as follows.
The Euler equation (\ref{nAbEu1}) can be rewritten as 
(no sum over $f$)
\be
j^\mu_f\Omega_{\mu\nu}^f=0\;,\;\; \Omega^f=d\Pi_f\;,
\ee
where $j^ \mu _f =\rho_f v_f^ \mu$, 
and by the same arguments as in the Abelian case in \cite{Abanov:2021hio}, 
the helicity ${\cal H}'_{\rm tot}$ is conserved:
\be
dh_f=\Tr[\Omega_f\wedge \Omega_f]\;,
\ee
but then by $j_f^ \mu \Omega_{\mu\nu}^ f=0$ (no sum over $f$), 
we have $\Omega_f\wedge \Omega_f=0$, which implies that $dh_f=0$, so 
${\cal H}' _f=\int h_f$ is conserved. But {\em there is no interaction between 
the various $f$'s, so this is somewhat trivial}.

\section{Conclusions}

In this paper we have constructed non-Abelian helicities for non-Abelian fluids, first by themselves, and then
coupled to Yang-Mills fields. 
One type of fluid was the "gluonic" fluid, in the adjoint, with densities $\rho^a$ and 
velocities $u^{\mu a}$, for which we have 
defined the total helicities ${{\cal H}^{NA}}^a_{mm}$ and ${\cal H}^{NA}_{\rm tot}$, 
and the other was the "quark" fluid, 
with densities $\rho_f$ and velocities $u^\mu_f$, with $f$ in the Cartan subalgebra 
of $U(N)$, for which we have defined 
the helicities ${\cal H}'_{\rm tot}$ and ${\cal H}''_{\rm tot}$.

We then have constructed non-Abelian Euler equations that conserve the above 
helicities. In the gluonic case, we have 
obtained Euler equations with an index $a$, in which the current $j_\mu^a$ does 
not factorize into $j_\mu$ and $Q^a$, 
as in previous attempts. We have also found a Lagrangian that gives these equations, 
and constructed classical solutions. 
In the quark case, the non-Abelian Euler equation we have found conserves the corresponding 
helicity, but the fluid types 
$f$ do not interact. 

In our construction, we have reviewed the Abelian Lagrangian construction for non-relativistic and 
relativistic fluids, using a Clebsch parametrization, coupled to electromagnetism,
and written also the symmetries, and Hamiltonian descriptions, which were perhaps not 
previously done. Following that, we have constructed two direct non-Abelian analogs of the Abelian Clebsch 
construction, but we have found that for one, the resulting Euler-like equation involved also the Clebsch fields, 
and did not conserve the helicities, so did not serve our purposes, while for the other we had a single 
equation (not a matrix, or $N^2$ components, equation). But perhaps these can be thought of as a 
good mathematical physics problems, that can have other applications. 

There are many things left for further work, like for instance finding 
classical solutions in the presence of dynamical Yang-Mills 
fields, which we have not done. Also, the generalization to higher 
knottedness of the solutions is still left for the future.

\section*{Acknowledgments}

We would like to thank A. Abanov  and P. Wiegmann 
for useful discussions. 
The work of HN is supported in part by  CNPq grant 304583/2023-5
and FAPESP grant 2019/21281-4.
HN would also like to thank the ICTP-SAIFR for their support through FAPESP grant grant 2021/14335-0.
The work of JS was supported  by a grant 01034816 titled  ``String theory reloaded- from fundamental 
questions to applications''     
of  the ``Israel Planning and budgeting committee'' (Vatat). 

\appendix

\section{The fluid Hopfion solution}

In \cite{Alves:2017ggb,Alves:2017zjt}, 
by mapping the electromagnetic Hopfion solution, the fluid Hopfion was derived. 
The velocity components take the form
\bea
v_x&=&\frac{2(y+x(t-z))}{1+x^2+y^2+(t-z)^2},\cr 
v_y&=&\frac{-2(x-y(t-z))}{1+x^2+y^2+(t-z)^2},\cr
v_z^2&=&1-v_x^2-v_y^2.\label{velhopf}
\eea

\begin{figure}[t!]
\begin{tabular} {c}
\includegraphics[width=7cm]{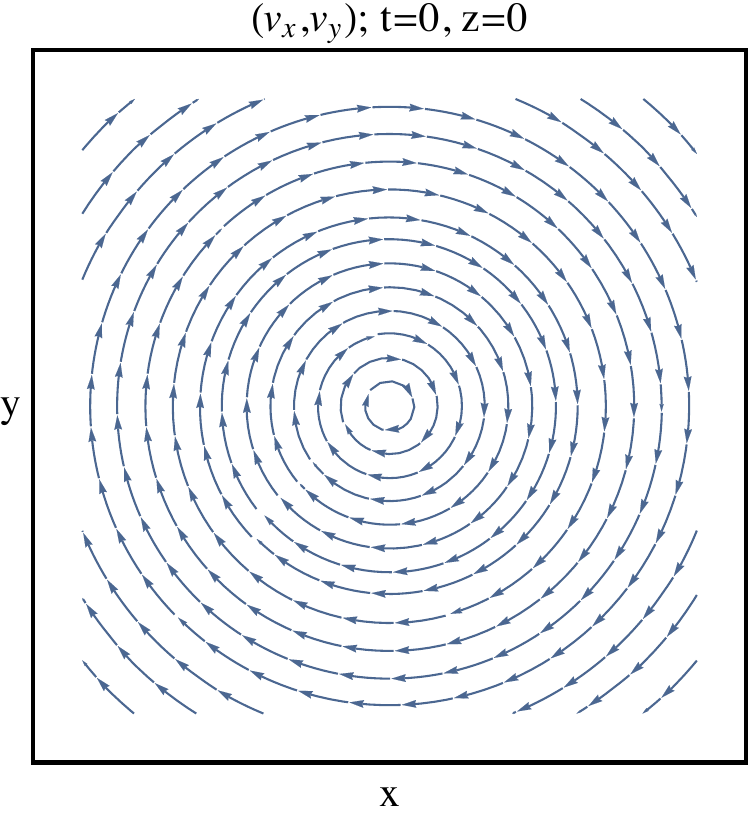}  \includegraphics[width=7cm]{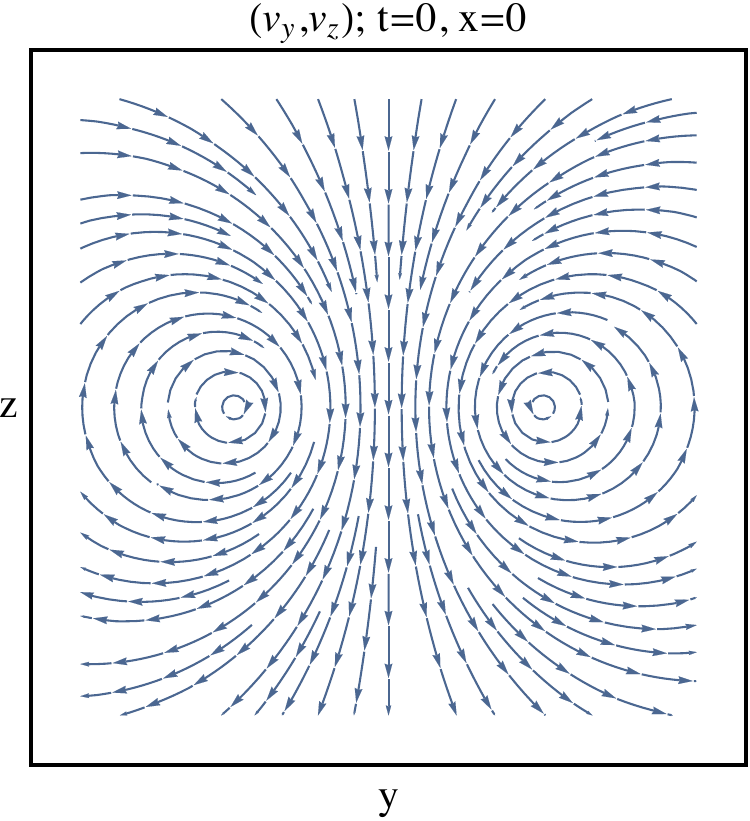}
\end{tabular}
\caption{Orthogonal sections of the velocity field for the Hopfion solution, on the $(x,y)$ plane 
(left) and $(y,z)$ plane (right).  
Using rotational symmetry in the $(x,y)$ directions the linked torus structure is apparent.}
\label{fighopf}
\end{figure}
\begin{figure}[t!]
\begin{tabular} {c}
\includegraphics[width=7cm]{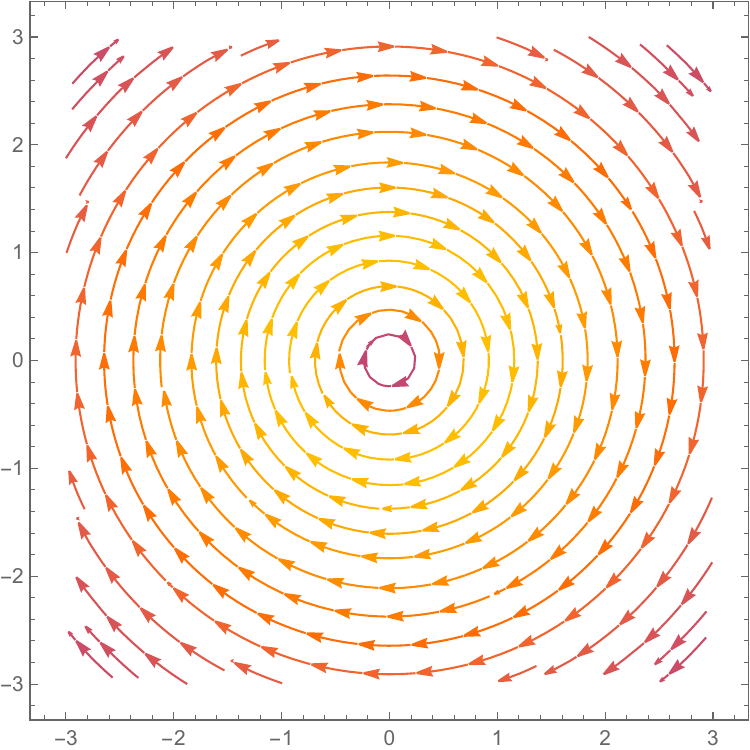}  \includegraphics[width=7cm]{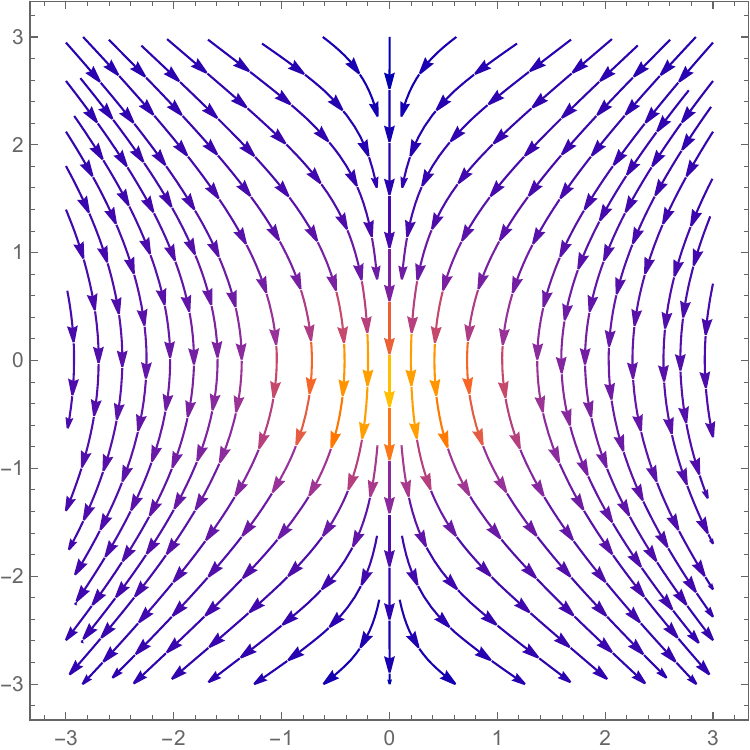}
\end{tabular}
\caption{Orthogonal sections of the vorticity field for the Hopfion solution, $(\omega_x,\omega_y)$ 
on the $(x,y)$ plane (left) and  $(\omega_y,\omega_z)$$(y,z)$ plane (right). } 
 \label{fighopf}
\end{figure}

The vorticity components are given by 
\be
\omega_x= \frac{2 \left(t^2 y-2 z (t y+x)+2 t x+x^2 y+y^3+y z^2+y\right)}{\left((t-z)^2+x^2+y^2+1\right)}\;,
\ee
\be
\omega_y=-\frac{2 \left(x \left((t-z)^2+y^2+1\right)+2 y
   (z-t)+x^3\right)}{\left((t-z)^2+x^2+y^2+1\right)^2}\;,
	\ee
	\be
\omega_z=	-\frac{4 \left((t-z)^2+1\right)}{\left((t-z)^2+x^2+y^2+1\right)^2}.
	\ee
It is easy to check that $\vec\nabla\cdot\vec \omega=0$. 
The scalar product of $\vec v$ and $\vec \omega$ is given by
\be
\vec v\cdot\vec \omega = \frac{4 \left((t-z)^2+x^2+y^2-1\right)}{\left((t-z)^2+x^2+y^2+1\right)^2}.
\ee

\bibliography{Nonabelianfluid}

\bibliographystyle{utphys}

\end{document}